\documentclass[preprint,3p]{elsarticle}

\usepackage{graphicx}
\usepackage{amssymb}
\usepackage[abs]{overpic} 
\usepackage{color}
\usepackage{booktabs}

\journal{Nuclear Physics A}

\begin{document}

\begin{frontmatter}

\title{Hypernuclear spectroscopy of products from $^6$Li projectiles on~a~carbon~target~at~2~$A$~GeV}

\author[guissen,gsi]{C. Rappold\corref{cor1}}
\ead{c.rappold@gsi.de}
\cortext[cor1]{Corresponding author}

\address[guissen]{Justus-Liebig-Universit\"at Giessen, Heinrich-Buff-Ring 16, 35392 Giessen, Germany}
\address[gsi]{GSI Helmholtz Centre for Heavy Ion Research, Planckstrasse 1, 64291 Darmstadt, Germany}

\author[gsi,snu]{ E. Kim}
\address[snu]{Seoul National University, Gwanakro Sillim-dong, Gwanak-gu, Seoul 151-747, Republic of Korea}

\author[gsi,ut]{ D. Nakajima}
\address[ut]{The University of Tokyo, 7-3-1 Hongo, Bunkyo-ku, Tokyo 113-0033, Japan}

\author[gsi,mainz,him]{T.R. Saito\corref{cor1}} 
\ead{t.saito@gsi.de}

\address[mainz]{Johannes Gutenberg-Universit\"at Mainz, J.J.Becherweg 40, 55099 Mainz, Germany}
\address[him]{The Helmholtz Institute Mainz (HIM), J.J.Becherweg 40, 55099 Mainz, Germany}

\author[gsi,mainz]{ O. Bertini}

\author[gsi]{ S. Bianchin}

\author[gsi,nut]{V. Bozkurt}
\address[nut]{Nigde University, 51100 Nigde, Turkey}

\author[kvi]{ M. Kavatsyuk}
\address[kvi]{KVI, University of Groningen, Zernikelaan 25, NL-9747 AA Groningen, The Netherlands}

\author[gsi,mainz]{ Y. Ma}

\author[gsi,mainz,him]{ F. Maas}

\author[gsi]{ S. Minami}

\author[gsi]{ B. \"Ozel-Tashenov}

\author[gsi,him,ouj]{ K. Yoshida} 

\address[ouj]{Osaka University, 1-1 Machikaneyama, Toyonaka, Osaka 560-0043, Japan}

\author[mainz]{ P. Achenbach}

\author[rcnp]{ S. Ajimura}
\address[rcnp]{Research Centre for Nuclear Physics (RCNP), 10-1 Mihogaoka, Ibaraki, Osaka 567-0047, Japan}

\author[tud,gsi]{ T. Aumann}
\address[tud]{Technische Universit\"{a}t Darmstadt, 64289 Darmstadt, Germany}

\author[mainz]{ C. Ayerbe Gayoso}

\author[snu]{ H.C. Bhang}

\author[tud]{ C. Caesar} 

\author[nut]{ S. Erturk}

\author[oecu]{ T. Fukuda} 
\address[oecu]{Osaka Electro-Communication University, Hatsu-cho 18-8, Neyagawa, Osaka 572-8530, Japan}

\author[gsi,nut]{ B. G\"ok\"uz\"um}

\author[kvi]{ E. Guliev}

\author[kuj]{ T. Hiraiwa}
\address[kuj]{Kyoto University, Kyoto 606-8502, Japan}

\author[gsi]{ J. Hoffmann}

\author[gsi]{ G. Ickert}

\author[nut]{ Z.S. Ketenci}

\author[gsi,mainz]{ D. Khaneft}

\author[snu]{ M. Kim} 

\author[snu]{ S. Kim} 

\author[gsi]{ K. Koch} 

\author[gsi]{ N. Kurz}

\author[gsi,subatech]{A.~Le~F\`evre}
\address[subatech]{SUBATECH, La Chantrerie, 4 rue Alfred Kastler, BP 20722, 44307 Nantes cedex 3, France}

\author[oecu]{ Y. Mizoi}

\author[kuj]{ M. Moritsu}

\author[kuj]{ T. Nagae}

\author[mainz]{ L. Nungesser}

\author[kuj]{ A. Okamura}

\author[gsi]{ W. Ott} 

\author[mainz]{ J. Pochodzalla}

\author[ouj]{ A. Sakaguchi}

\author[kuj]{ M. Sako}

\author[gsi]{ C.J. Schmidt}

\author[kek]{ M. Sekimoto}
\address[kek]{KEK, 1-1 Oho, Tsukuba, Ibaraki 305-0801, Japan}

\author[gsi]{ H. Simon}

\author[kuj]{ H. Sugimura}

\author[kek]{ T. Takahashi}

\author[kvi]{ G.J. Tambave}

\author[tuj]{ H. Tamura} 
\address[tuj]{Tohoku University, 6-3 Aoba Aramaki Aoba Sendai, Miyagi 980-7875, Japan}

\author[gsi]{ W. Trautmann}

\author[gsi]{ S. Voltz} 

\author[kuj]{ N. Yokota} 

\author[snu]{ C.J. Yoon}

\date{\today}

\begin{abstract}

A novel experiment, aiming at demonstrating the feasibility of hypernuclear spectroscopy with heavy-ion beams, was conducted. Using the invariant mass method, the spectroscopy of hypernuclear products of  $^6$Li projectiles on a carbon target at 2 $A$ GeV was performed. Signals of the $\Lambda $-hyperon and $_\Lambda^3$H and $_\Lambda^4$H hypernuclei were observed for final states of p+$\pi ^-$, $^3$He+$\pi ^-$ and $^4$He+$\pi ^-$, respectively, with significance values of 6.7, 4.7 and 4.9$\sigma $. By analyzing the proper decay time from secondary vertex distribution with the unbinned maximum likelihood fitting method, their lifetime values were deduced to be $262 ^{+56}_{-43} \pm 45$ ps for $\Lambda $, $183 ^{+42}_{-32} \pm 37$ ps for $_\Lambda^3$H, and $140 ^{+48}_{-33}\pm 35 $ ps for $_\Lambda^4$H.  
\end{abstract}

\begin{keyword}
hypernuclear spectroscopy \sep heavy ion induced reaction \sep invariant mass \sep lifetime measurement

\end{keyword}

\end{frontmatter}

\section{Introduction}
One of the key issues in nuclear and hadron physics is the investigation of baryon-baryon interactions under the flavor SU(3) symmetry with up, down and strange quarks. A baryon involving at least one strange quark is called a hyperon, and the lightest hyperon is $\Lambda$. Because the $\Lambda $-hyperon decays via weak interaction with a lifetime of 263.2 ps \cite{cite:pdg}, it has not been practical to study the nucleon-$\Lambda $ and $\Lambda$-$\Lambda$ interactions by direct reaction experiments with projectiles and targets involving the $\Lambda $-hyperon. Therefore a $\Lambda$-hypernucleus, a bound subatomic system containing a $\Lambda$, was studied to extract information on the nucleon-$\Lambda$ and $\Lambda$-$\Lambda$ interactions. Lambda-hypernuclei until now have mainly been investigated experimentally with emulsion techniques \cite{Davis20053}, secondary meson beams \cite{Hashimoto2006564} and primary electron beams \cite{jlab}. In those experiments, the isospin of the produced hypernuclei was limited by the reaction mechanism since a nucleus in the stable target material is converted to a $\Lambda$-hypernucleus by production or exchange of strangeness in a single nucleon. With heavy ion beams, a central collision of platinum projectiles at 11.5 GeV/c on a gold target was used successfully to produce and identify $_\Lambda^3$H (hypertriton) and $_\Lambda^4$H hypernuclei \cite{ags}. Recently, the STAR collaboration used relativistic heavy ion collisions (Au+Au) to study hypertriton and anti-hypertriton \cite{cite:STAR}. However, the latter two experiments to date have not proven suitable for the production of hypernuclei with the mass value $\mathrm{A}\geq4$. 

A different approach to studying hypernuclei by using projectile fragmentation reactions of heavy ion beams was employed for the present work. In such reactions, a projectile fragment can capture a hyperon produced in the hot participant region to produce a hypernucleus. In this reaction, the energy of heavy ion beams should exceed the energy threshold for the hyperon production, and the velocity of the produced hypernucleus should be similar to that of the projectile. Thus, the produced hypernucleus has a large Lorentz factor, and the decay of the hypernucleus takes place well behind the production target. This makes it possible to study hypernuclei in flight. Since a hypernucleus is produced from a projectile fragment, isospin and mass values of the produced hypernuclei, unlike in other hypernuclear experiments, can be widely distributed in similar fashion. The first attempt was made with $^{16}$O beams at 2.1 $A$ GeV on a polyethylene target \cite{lbl}. However, the measured cross section of the hypernuclear candidates greatly exceeded the theoretical expectations, which was ascribed to a difficulty with implementing the kaon trigger. Subsequently, an experiment performed by bombarding a polyethylene target with beams of $^4$He at 3.7~$A$~GeV and $^7$Li at 3.0~$A$ ~GeV was successful in producing light hypernuclei \cite{dubna1,dubna2}. However, since the invariant mass of the final states was not measured, there was ambiguity in identifying the hypernuclei produced. 

As mentioned earlier, one of the unique features of the hypernuclear spectroscopy with projectile fragmentation reactions is that due to a large Lorentz factor of the produced hypernuclei the decay of the hypernuclei can be observed in flight behind the production target. Lifetime values of observed hypernuclei can be deduced by studying the distribution of the \emph{proper decay time}, obtained from the measured decay length, in the rest frame of the mother state of interest, thus making the deduced lifetime values independent of the detectors' time resolution. The lifetimes of hypernuclei are of interest since they are sensitive to the overall wave function of the hyperon located within the core nucleus. The lifetime of light $\Lambda$-hypernuclei has been conjectured to be similar to the lifetime of a free $\Lambda$ hyperon if it is weakly bound to the core nucleus \cite{PhysRevC.57.1595}. Deviations from the value of 263.2 ps of the $\Lambda$ lifetime \cite{cite:pdg} would possibly provide new information on Lambda's wavefunction inside hypernuclei. 
In early days, lifetime measurements for hypernuclei were derived from early emulsion experiments \cite{cite:emulsion1,cite:emulsion2,cite:emulsion3,cite:emulsion4,cite:emulsion5} and bubble chamber experiments \cite{cite:bubble1,cite:bubble2,cite:bubble3}. Subsequently, there were a few experiments with a missing-mass method to extract the lifetime values of light hypernuclei  \cite{cite:counter1,cite:counter2}. Lifetime values of heavier hypernuclei were also measured by observing the nonmesonic weak decay channel in coincidence with the $(K^-,\ \pi^-)$ and $(\pi^+,\ K^+)$ reactions \cite{cite:NMWD1,cite:NMWD2,cite:NMWD3,cite:NMWD4}.  
Recently, the lifetime of the hypertriton and anti-hypertriton were measured at the Relativistic Heavy Ion Collision \cite{cite:STAR}, in a way that was similar to the present experiment.
However, for most of those experiments, the reported one-sigma interval of measurements exhibits large discrepancies.

The HypHI collaboration has proposed a series of experiments at the GSI Helmholtz Centre for Heavy Ion Research, using induced reactions of stable heavy ion beams and rare-isotope beams, with the aim of producing and measuring hypernuclei with an invariant mass method \cite{LOI}. In the proposed experiments, charged particles and neutrons from the mesonic or non-mesonic weak decay of hypernuclei are tracked and identified in order to reconstruct the hypernuclear mass values. The lifetime of produced hypernuclei can be extracted by measuring of the proper time in the rest frame of the hypernuclear decay. This methodology also allows investigating neutron- and proton-rich hypernuclei as well as hypernuclei with more than two units of strangeness, and several hypernuclei can be studied in a single data taking.

A first experiment to study $^3_\Lambda $H and $^4_\Lambda $H hypernuclei was performed by mean of projectile fragmentation reactions of $^6$Li projectiles at 2 $A$ GeV delivered on a carbon target. Results of the observed invariant mass distributions of p+$\pi ^-$ for $\Lambda $, $^3$He+$\pi ^-$ for $^3_\Lambda $H and  $^4$He+$\pi ^-$ for $^4_\Lambda $H as well as their estimated lifetime values will be discussed in this article.

\begin{figure*}[htb]
\centering
  \resizebox{150mm}{!}{\includegraphics{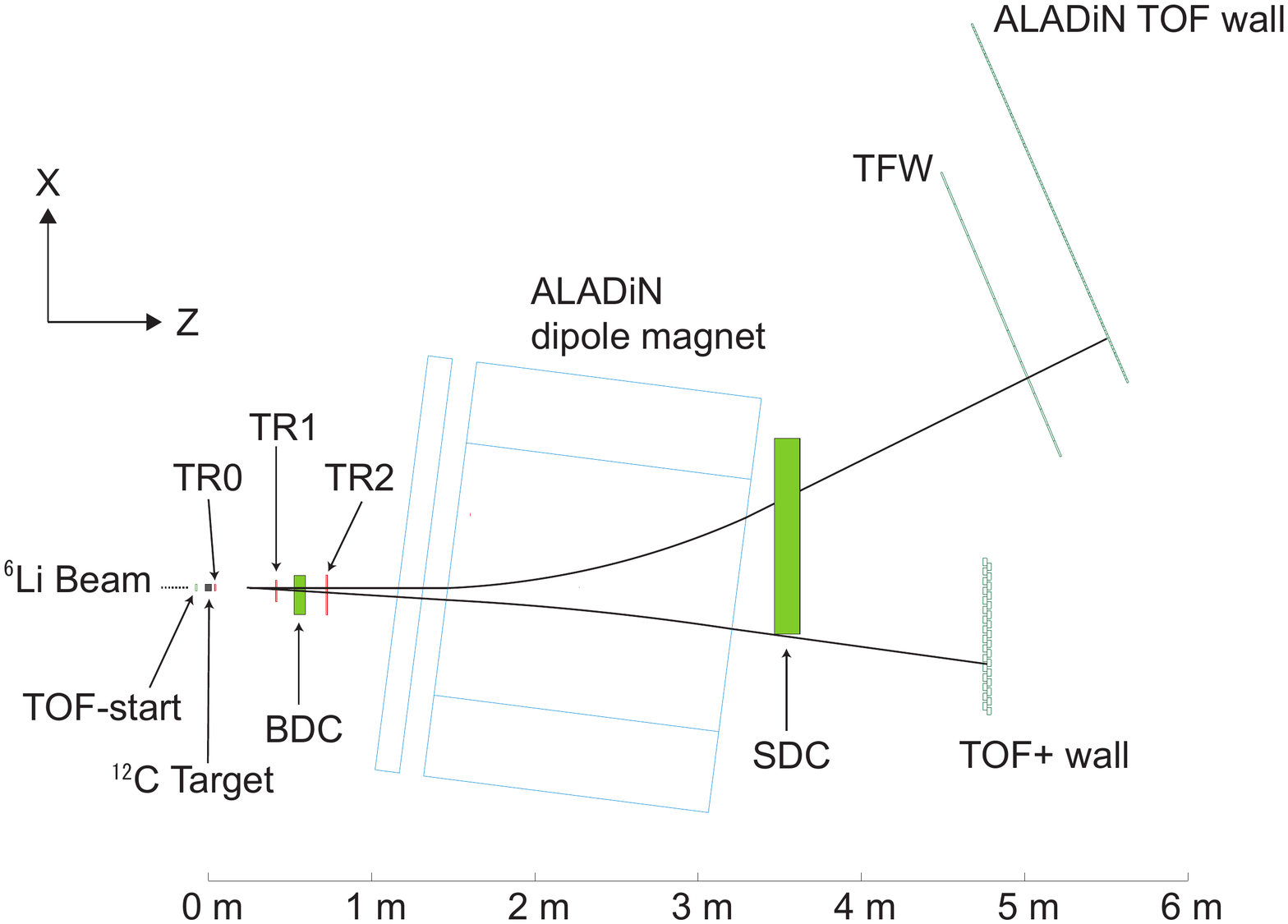}} 
\caption{Schematic layout of the HypHI Phase 0 experimental setup.}
\label{fig:setup}

\end{figure*}

\section{HypHI Phase 0 experiment}
\label{sec:experiment}

The first HypHI experiment, Phase 0,  took place at the GSI Helmholtz Centre for Heavy Ion Research. Although the experimental setup of the Phase 0 experiment has already been presented in Ref. \cite{cite:torino,hyphinpa1}, it will be briefly discussed below. 

Figure \ref{fig:setup} shows a schematic layout of the Phase 0 experiment. Projectiles of $^6$Li at 2 $A$ GeV with an average intensity of 3$\times 10^6$ beam particle per second were propelled at a carbon graphite target 8.84~g/cm$^2$ thick. The ALADiN magnet was used as a bending magnet for charged particles produced from the target and hypernuclear decay vertices, as shown in Figure \ref{fig:setup}. A magnetic field of approximately 0.75 T was applied. The distance between the target and the center of the ALADiN magnet was 2.35 m. 
A small array of plastic finger hodoscopes labeled in the figure as TOF-start was used as a start counter for Time-of-Flight (TOF) measurement. In order to track charged particles, three layers of scintillating fiber detectors \cite{fiber}, identified as TR0, TR1 and TR2 in the figure, were set in front of the ALADiN magnet. TR0 was placed at 4.5 cm behind the target, and the distance of TR1 and TR2 from the target center was 40 and 70 cm respectively. A drift chamber BDC with six wire layers ($xx'$, $uu'$ and $vv'$) to track charged particles was positioned between the two fiber detectors arrays, TR1 and TR2. Since it was installed around the beam axis, the sense wires were wrapped with Teflon to create a local beam killer. 
Behind the ALADiN magnet, two hodoscopes with plastic scintillating bars, TFW and ALADiN TOF wall, provided the stop signal for TOF measurements and the hit position information for $\pi ^-$ mesons. 
Positively charged particles and fragments were measured by another plastic hodoscope, labeled as TOF+ wall in the figure. An additional drift-chamber, labeled SDC, with four wire layers ($xx'$ and $yy'$), was positioned behind the ALADiN magnet to measure hit positions of outgoing charged particles. A beam killer was also made for SDC by disconnecting sense and potential wires to reduce the electric fields around the beam position. 

The trigger system for the data acquisition electronics combined three trigger stages. The first stage was a tracking trigger which is generated by signals from the scintillating fiber tracking arrays with VUPROM2 \cite{minami2_1,minami2_2}. It checked for a secondary vertex sites behind the target caused by free-$\Lambda $ and hypernuclear decays. The second stage was linked to $\pi ^-$ detection by the TFW wall. The third stage required the detection of $Z=2$ charged fragments in TOF+ wall. More details on the experimental setup and the preliminary results can be found in \cite{cite:torino}, while the final data analyses with a proper statistical methodology, based on the RooFit and RooStats frameworks included in the ROOT package \cite{cite:roofit,cite:roostats,cite:ROOT}, will be presented in this article. All detector information and the improved detector calibrations were included in the present analyses, contrary to Ref. \cite{cite:torino}. The measurement took place over a period of 3.5 days with an integrated luminosity of 0.066~pb$^{-1}$.

The tracking systems composed of scintillating fiber detector arrays and the two drift chambers were used for reconstructing tracks and determining the secondary vertex. The four scintillating hodoscope walls, used to perform time-of-flight measurements of charged particles, also worked as part of the tracking system. Track fitting was done by means of the Kalman filter algorithm, as summarized in \cite{cite:NimKalman}.

\section{Analysis: Particle identification, vertex selection and invariant mass}

First, the invariant mass distributions of the final states of two-body mesonic weak decays of  $\Lambda \to$~p~+~$\pi^-$, $_\Lambda^3$H~$\to ^3$He~+~$\pi^-$, $_\Lambda^4$H~$\to ^4$He~+~$\pi^-$ were analyzed. 

In each event, daughter candidates for those decays were identified and used to reconstruct the secondary vertex, as corresponding to the mesonic weak decay of interest. In conjunction with the invariant mass calculation, the mother bound state candidates were selected after a series of geometrical considerations in order to find the most probable candidate per event.

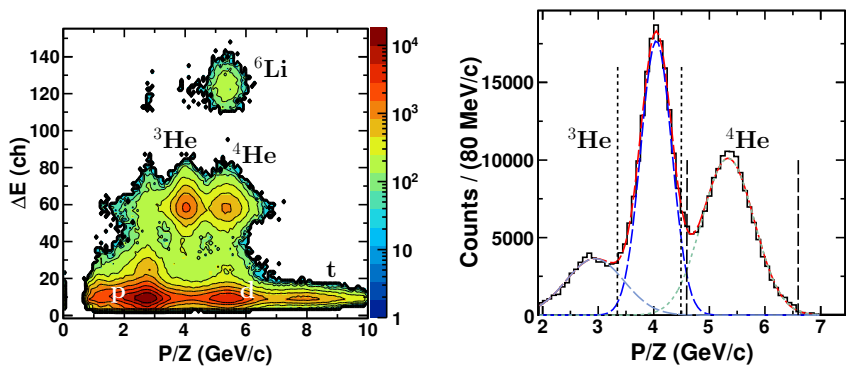
\begin{figure*}[htc]
\centering
\begin{tabular}{cc}

       \resizebox{55mm}{!}{
	\begin{overpic}[scale=0.5]%
      	  {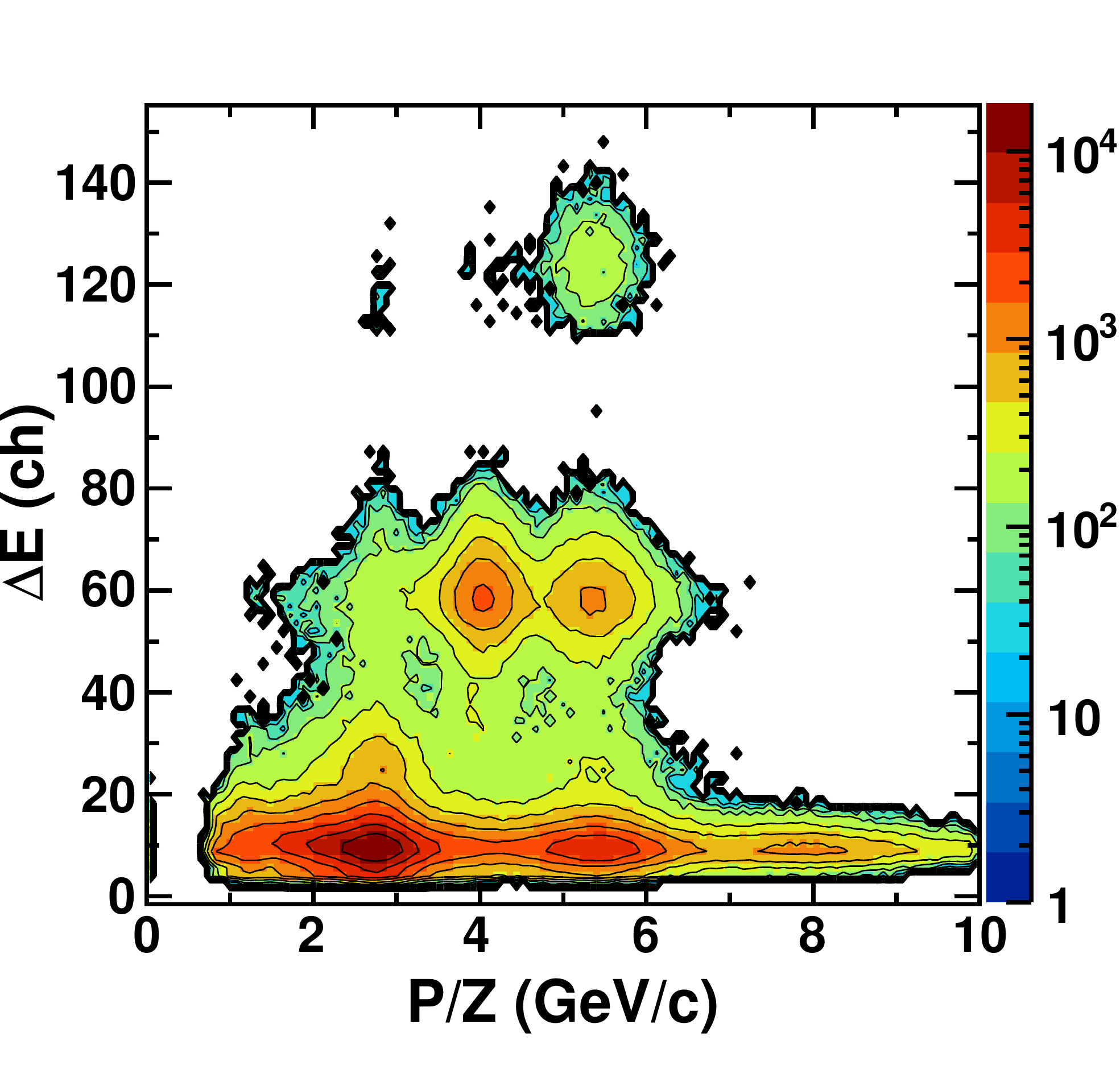}
	  \put(70,55){\textbf{\color{white}{\LARGE{p}}}}
	  \put(160,55){\textbf{\color{white}{\LARGE{d}}}}
	  \put(220,70){\textbf{\LARGE{t}}}
	  \put(100,160){\textbf{\LARGE{$^3$He}}}
	  \put(155,150){\textbf{\LARGE{$^4$He}}}
	  \put(170,210){\textbf{\LARGE{$^6$Li}}}
	\end{overpic}
	}

&
       \resizebox{55mm}{!}{
	\begin{overpic}[scale=0.5]%
      	  {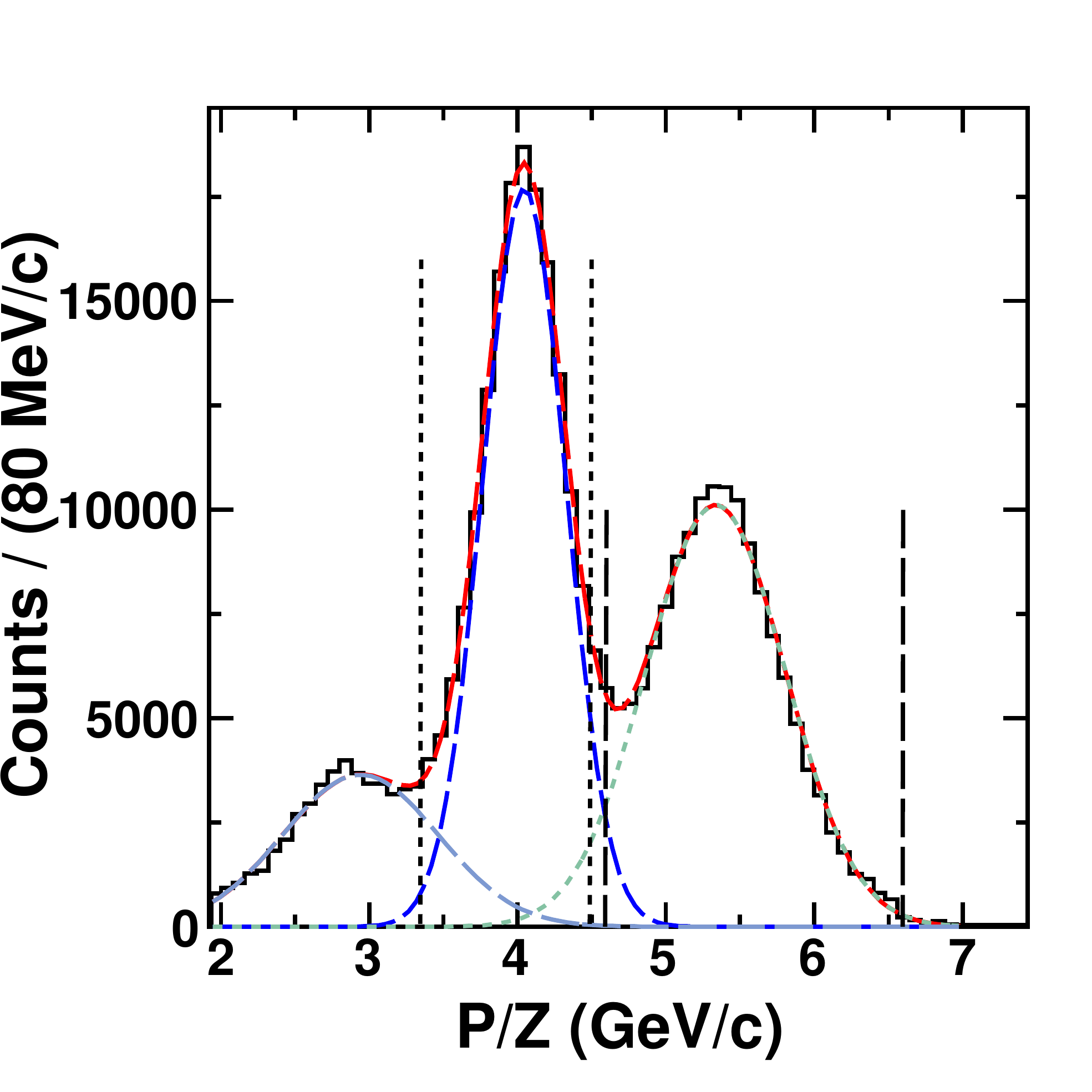}
	  \put(75,160){\textbf{\LARGE{$^3$He}}}
	  \put(185,160){\textbf{\LARGE{$^4$He}}}
	\end{overpic}
	} \\
	\end{tabular}
\caption{(Color online) Shown in the left panel is the correlation between the energy deposit of positively charged isotopes in the TOF+ wall detector and the momentum-to-charge ratio P/Z. The right panel depicts the projection of the momentum-to-charge ratio for the helium isotope identification in the right panel. The drawn-in vertical lines show the momentum intervals (3.35 GeV/c $\leq$ P/Z $\leq$ 4.5 GeV/c \& 4.6 GeV/c $\leq$ P/Z $\leq$ 6.6 GeV/c) used to distinguish $^3$He and $^4$He. Each contribution is modeled by a Gaussian probability density function shown individually by blue dashed lines and obtained from the combined fit of the distribution, shown as a red solid line.}
\label{fig:PID_mom_E}
\end{figure*}

After the track fitting procedure based on the Kalman Filter algorithm \cite{cite:NimKalman}, the different track candidates are associated with their p-values, the goodness-of-fit criteria. The p-value for each of the candidates is used to exclude poorly fitted tracks. The fitting procedure gives us the most probable momentum vector of the particle or fragment. The particle identification is obtained by applying the additional information from the hodoscope walls, such as the time-of-flight and the energy deposit. The charge, $Z$, of the positively charged particles and fragments can be determined from the correlation between the energy deposit, $\Delta E$, in the scintillating bars of the TOF+ wall and the momentum-to-charge ratio, $P/Z$, obtained by the track fitting.

The left panel of Figure \ref{fig:PID_mom_E} shows the correlation of $\Delta E\:\mathrm{vs.}\:P/Z$. A clear separation is seen between the hydrogen, helium and lithium isotopes. In the case of the He~isotopes, the species are determined by their momentum separation since they have a velocity close to that of the projectile: the time-of-flight measurement does not help in their identification. The right panel of Figure \ref{fig:PID_mom_E} exhibits the separation between the $^3$He and $^4$He species in the momentum-to-charge ratio distribution. Each species' contribution is modeled by a Gaussian probability density function and allows determining intervals of momentum-to-charge [3.35 GeV/c, 4.5 GeV/c] and [4.6 GeV/c, 6.6 GeV/c] for the identification of $^3$He and of $^4$He respectively. The $^3$He contribution to the contamination of the $^4$He identification is estimated to be approximately 1.7\%, while the contamination of $^4$He into the identification of $^3$He is approximately 1.8\%.

\begin{figure*}[htc]
\centering
\begin{tabular}{cc}

\resizebox{55mm}{!}{
	\begin{overpic}[scale=0.5]%
      	  {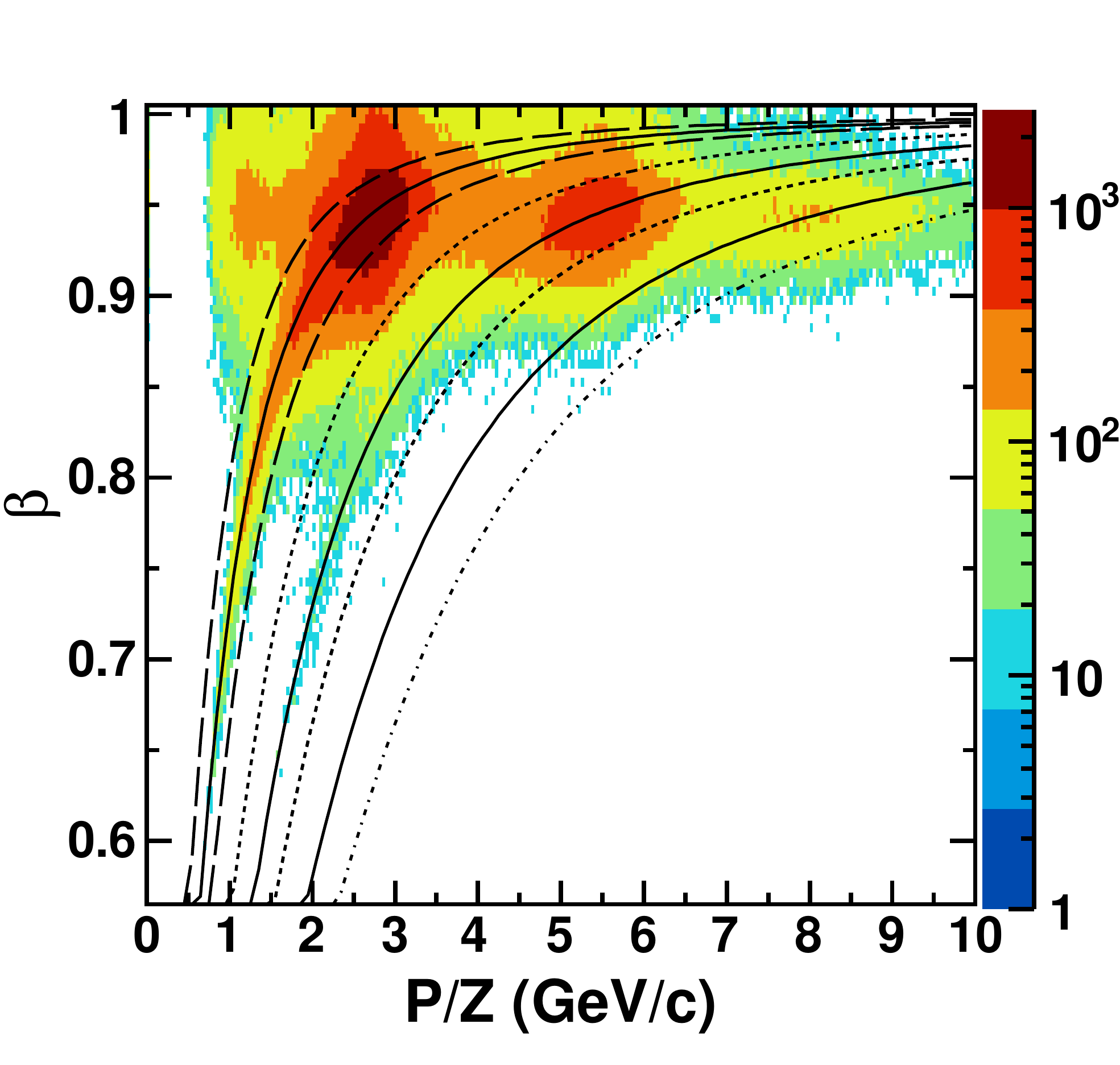}
	  \put(45,130){\textbf{\LARGE{p}}}
	  \put(80,110){\textbf{\LARGE{d}}}
	  \put(200,180){\textbf{\LARGE{t}}}
	  \end{overpic}
} &

\resizebox{55mm}{!}{
	\begin{overpic}[scale=0.5]%
      	  {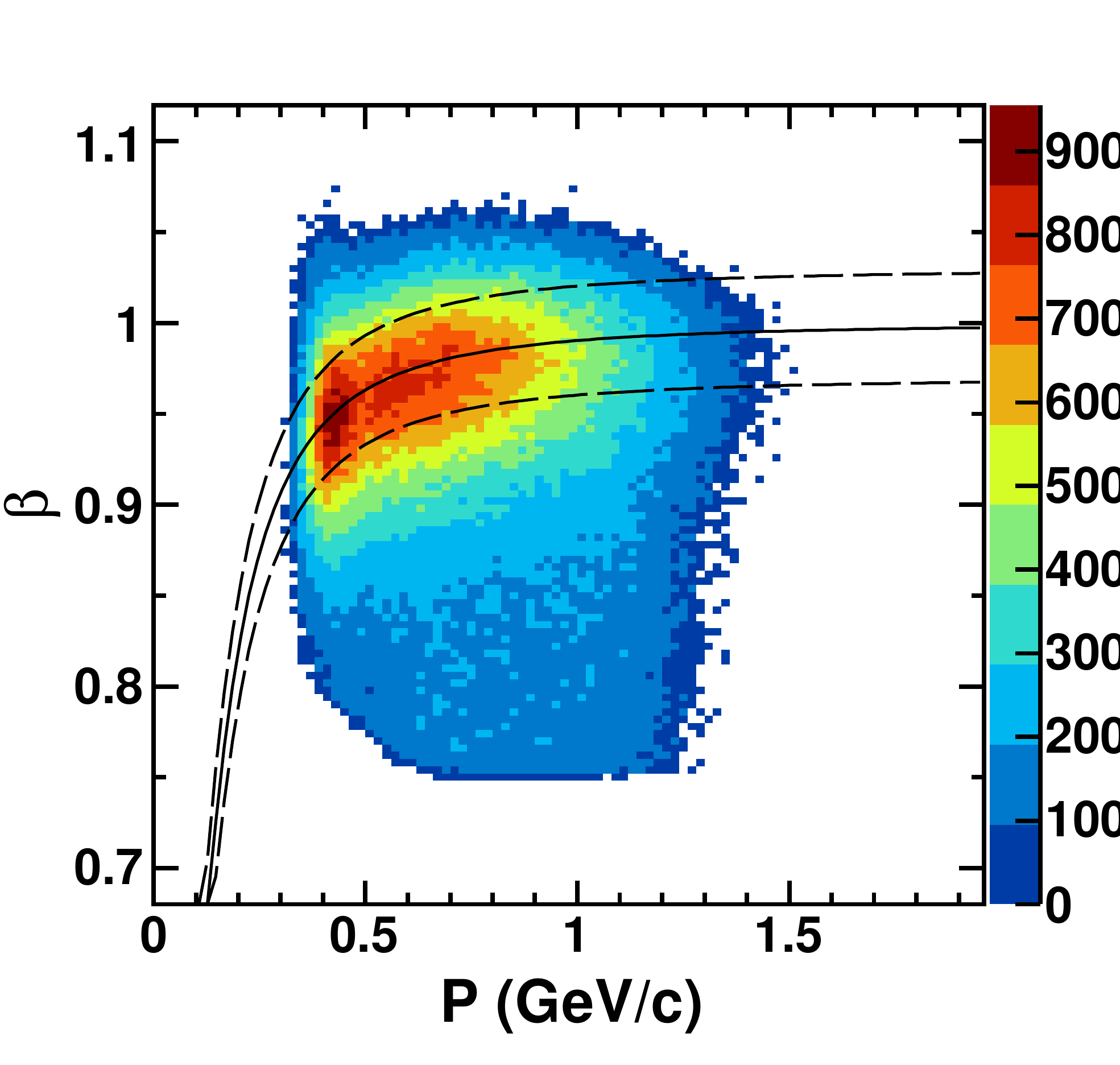}
	  \put(70,80){\textbf{\LARGE{${\pi^-}$}}}
	  \end{overpic}
	  }
\\

\end{tabular}
\caption{(Color online) Correlation between the velocity $\beta$ and the momentum for the Z=1 charge isotopes (left panel) and the $\pi^-$ meson (right panel). The theoretical function $\beta = 1/\sqrt{(m/p)^2+1}$ is drawn as a black line for each hydrogen isotope (proton, deuteron and triton) and the $\pi^-$ meson. Selection bands for the particle identification are drawn as dashed lines for each species of interest.}
\label{fig:PID_mom_beta}
\end{figure*}

In the case of hydrogen isotopes, the time-of-flight measurement is used to calculate the velocity, $\beta$, of each of the track candidates. The correlation of $\beta\:\mathrm{vs.}\:P/Z$ provides the identification of proton, deuteron and triton species. $\pi^-$ mesons are also identified with this correlation obtained from the time-of-flight measurement by the TFW detector. Figure \ref{fig:PID_mom_beta} shows those correlations for the positively charged particles (left panel) and the negatively charged particles (right panel). The species of hydrogen isotopes and $\pi^-$meson should be distributed around the theoretical function of $\beta = 1/\sqrt{(m/p)^2+1}$, where $m$ and $p$ respectively are the mass of the species of interest and the reconstructed momentum. Each theoretical line is indicated by a solid line in Figure \ref{fig:PID_mom_beta}. Each hydrogen species is identified by falling into the mass interval of $\pm$ 15\% from the nominal mass of the species of interest. In the left panel of Figure \ref{fig:PID_mom_beta}, the selection bands used for the identification are indicated by dashed lines, and can be observed in the correlation plot. The identification of $\pi ^-$ mesons is achieved in a manner similar to hydrogen isotopes by using the time-of-flight information at the TFW wall. The right panel of Figure \ref{fig:PID_mom_beta} shows the corresponding correlation of $\beta\:\mathrm{vs.}\:P/Z$, and the selection of $\pi ^-$ mesons is made by defining a band of $\pm\,0.05$ in $\beta$ around the theoretical line.

Once the particle and fragment identification has been performed, reconstruction of the secondary vertex is necessary in order to obtain the information on the $\Lambda$ hyperon and the hypernuclei of interest. The secondary vertex of interest corresponds to the two-body mesonic weak decay of the $\Lambda$ hyperon and of the $^3_\Lambda$H and $^4_\Lambda$H hypernuclei. First, the following pair of track candidates is associated to match that two-body decay: p+$\pi^-$, $^3$He+$\pi^-$ and  $^4$He+$\pi^-$. Next, several criteria are applied in selecting the best secondary vertex candidate per event among the vertex candidates of paired daughter particles. To begin with, the p-value from the track fitting for the daughter of interest has to be greater than 0.005 and 0.2 for the positively charged fragment and the $\pi^-$ meson candidates, respectively.

\begin{figure*}[htc]
\centering
\begin{tabular}{cc}

\resizebox{85mm}{!}{
	\begin{overpic}[scale=0.5]%
      	  {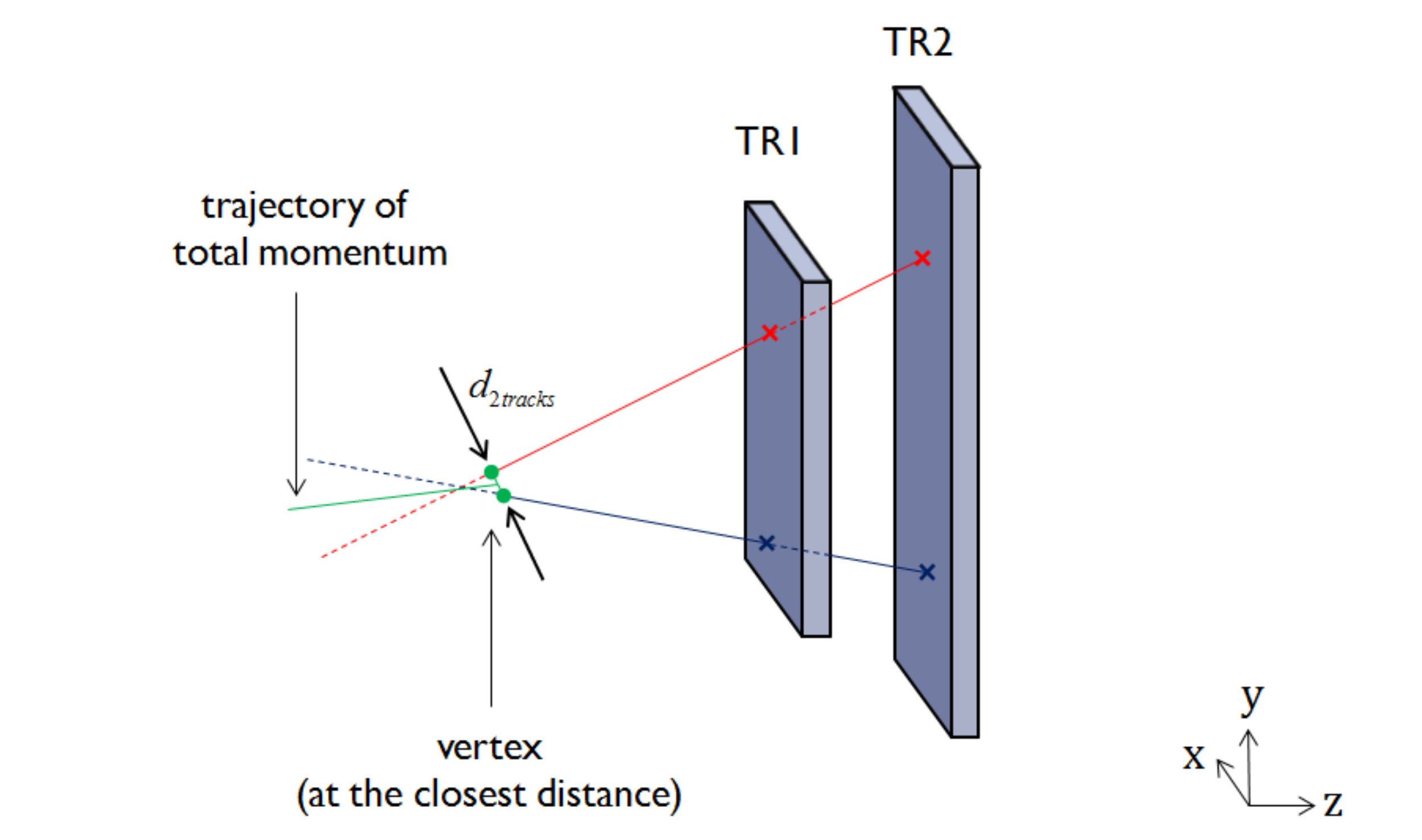}
	  \end{overpic}
} &
\resizebox{85mm}{!}{
	\begin{overpic}[scale=0.5]%
      	  {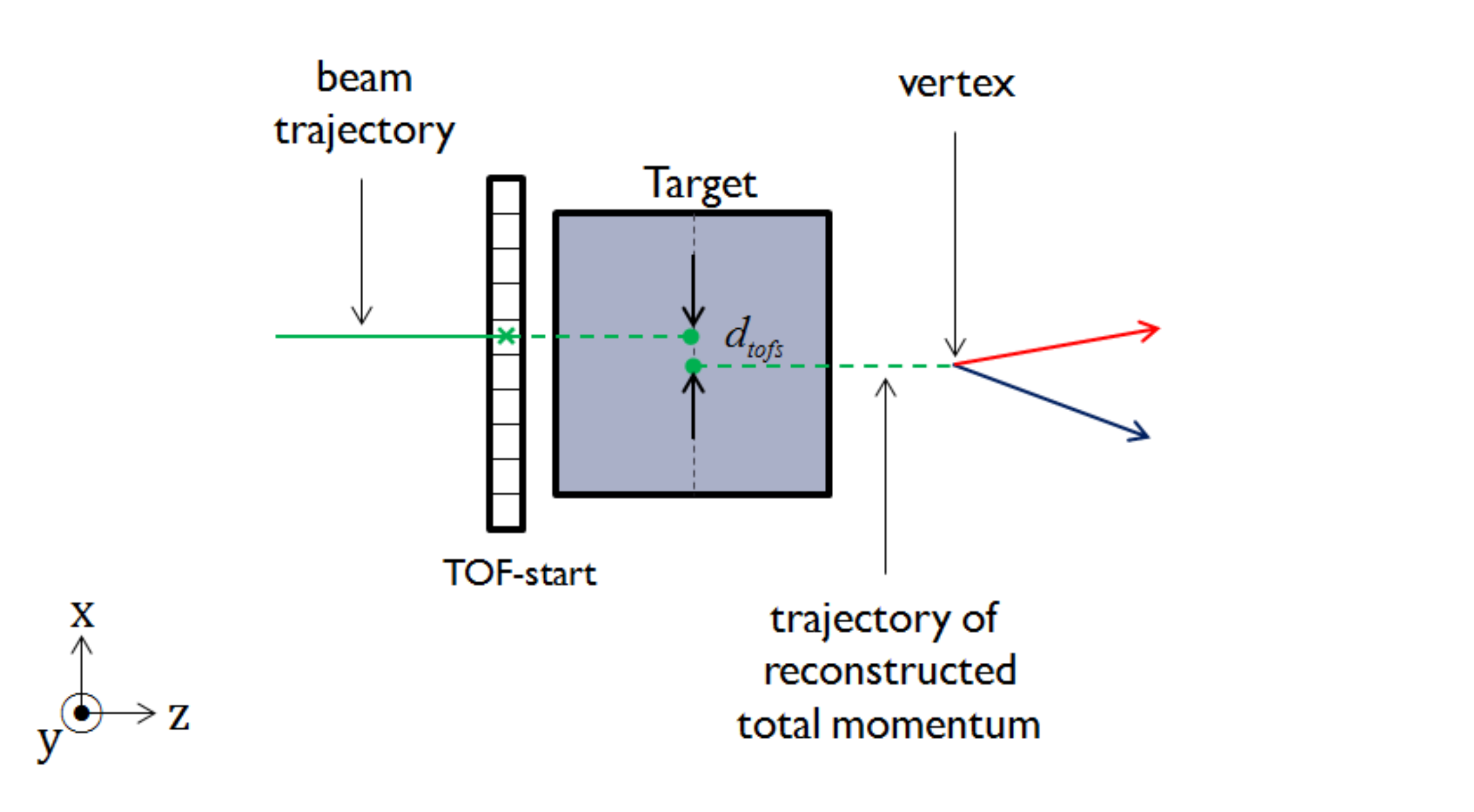}
	  \end{overpic}
	  }

\\

\end{tabular}
\caption{Geometrical rules for vertex selection. The distance of closest approach of the track pair and estimation of the secondary vertex position are shown in the left panel. The right panel shows the distance between the extrapolated position of the reconstructed mother track candidate and the beam position as measured by the TOF-start detector at the target position in the horizontal plane of the laboratory frame.}
\label{fig:vtx_selection}
\end{figure*}

Next, geometrical considerations are applied. The distance of closest approach for the track pair, $d_{2tracks}$, shown schematically in the left panel of Figure \ref{fig:vtx_selection}, is calculated to define an estimated the secondary vertex position. This distance has to be less than 4 mm for the secondary vertex candidate to be accepted. This is followed by evaluating the distance in the horizontal plane of the laboratory frame between the extrapolated position of the reconstructed mother track candidate at the target position and the beam position measured by the TOF-start detector. A schematic view of this distance $d_{tofs}$ is shown in the right panel of Figure \ref{fig:vtx_selection}. During the experiment, the beam trajectories were adjusted by the accelerator to be perpendicular to the surface layer of the TOF-start detector, allowing the hit position of the TOF-start detector to represent the position of the production vertex in the target.

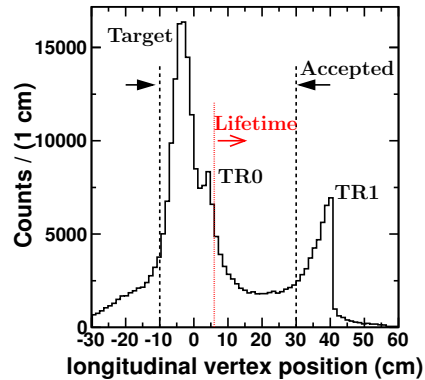
\begin{figure*}[htc]
\centering
\resizebox{55mm}{!}{
	\begin{overpic}[scale=0.5]%
      	  {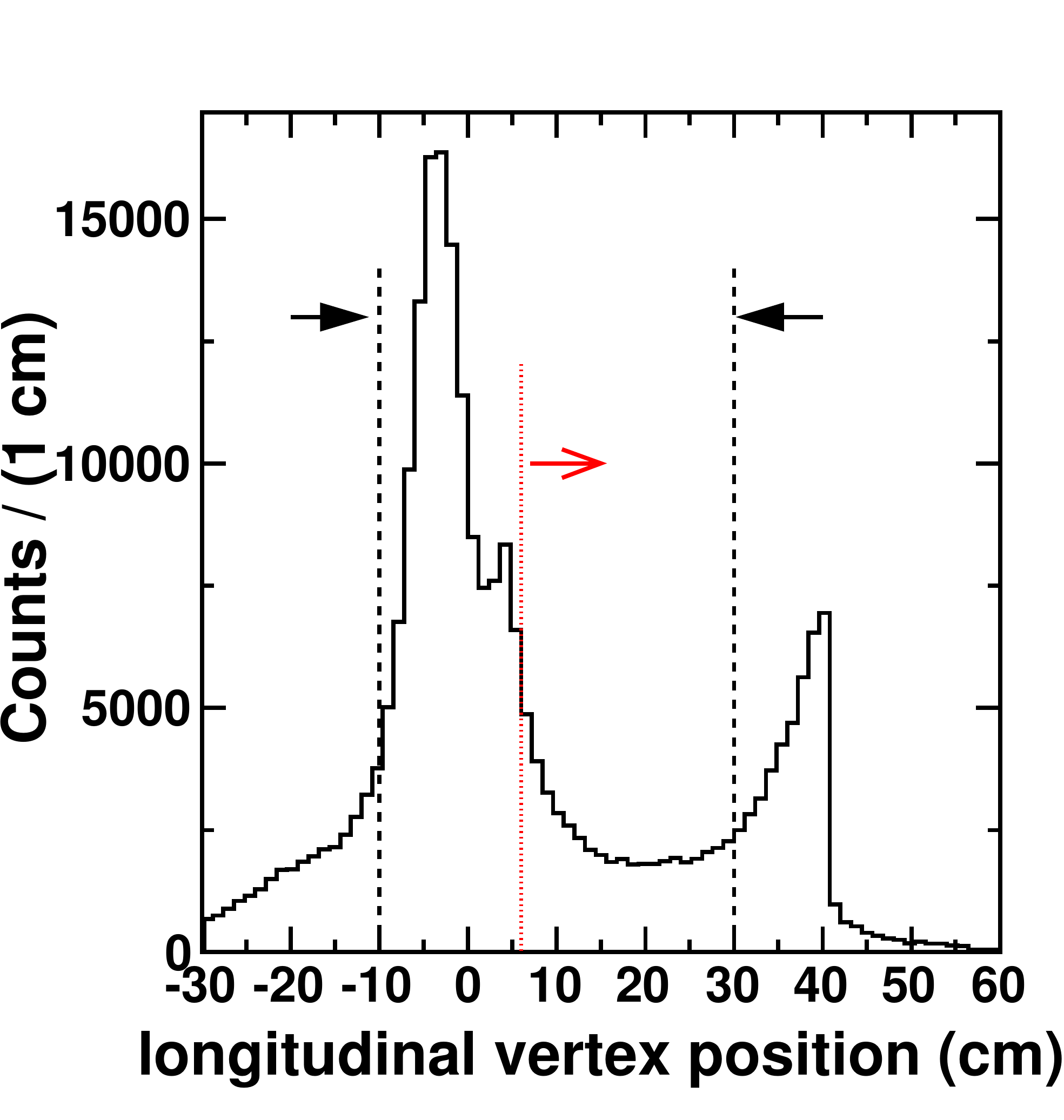}
 		  \put(223,135){\textcolor{black}{\textbf{\Large{TR1}}}}
 		  \put(65,245){\textcolor{black}{\textbf{\Large{Target}}}}
 		  \put(142,146){\textcolor{black}{\textbf{\Large{TR0}}}}
 		  \put(199,222){\textcolor{black}{\textbf{\Large{Accepted}}}}
 		  \put(139,183){\textcolor{red}{\textbf{\Large{Lifetime}}}}

	  \end{overpic}
}

\caption{Longitudinal secondary vertex position in the laboratory frame. The tallest dashed black vertical lines correspond to the interval selection for the secondary vertex candidate vertex position (-10 cm $\leq$ Z vertex position $\leq$ 30 cm). The shorter red vertical line corresponds to the requirement for the lifetime measurements ( Z vertex position $>$ 6 cm).}
\label{fig:vtx_Z}
\end{figure*}

The last rule applied in the vertex selection is based on the calculated longitudinal vertex position of the secondary vertex of interest in the laboratory frame. The secondary vertex is expected to be between the location of the production target and the first layer of the TR1 fiber detector. Figure \ref{fig:vtx_Z} shows the distribution of the longitudinal Z positions of all secondary vertex candidates. The contribution from the target and the TR0 fiber detector can be seen at -4 cm and 2 cm, respectively. The contribution from combinatorial background becomes dominant after 30 cm, peaking at the TR1 fiber detector position. %
The accepted secondary vertex candidate must have a longitudinal vertex position between -10 cm and 30 cm in the laboratory frame. 

After applying the rules for vertex selection, the invariant mass is calculated from the reconstructed four-vector of the pair of daughter candidates, (p, $\pi^-$), ($^3$He, $\pi^-$) and ($^4$He, $\pi^-$). For the estimation of the lifetime of the mother states of interest, the longitudinal position of the secondary vertex must be greater than 6 cm at a minimum.   

\begin{table*}[tHb] %
\centering
\caption{\label{tab:mass_calibration} The mean values of the invariant mass of $\Lambda$, $^3_{\Lambda}$H and $^4_{\Lambda}$H used for the calibration. A linear fit was used as calibration. 
 }

\begin{tabular}{cccc}
\toprule
 & $\Lambda$  & $^3_{\Lambda}$H & $^4_{\Lambda}$H \\
\midrule
Mass before calibration (Mean Value) (GeV) & 1.10785 & 2.9848 & 3.9048 \\
Mass of references (GeV)& 1.115683 & 2.99114 &  3.9225 \\

\bottomrule

\end{tabular}
\end{table*}

Due to slight uncertainty concerning the strength of the magnetic field during the experiment as well as the absolute position of one of the tracking detectors (SDC), the peak positions of the invariant mass distributions deviated from the known mass values, requiring calibration of the mass values. Table \ref{tab:mass_calibration} summarizes the values used for the calibration. A linear fit between the obtained values from the invariant mass distribution without calibration and the reference value of the masses of $\Lambda$, $^3_{\Lambda}$H and $^4_{\Lambda}$H was used.

\begin{figure*}[thb]
\centering
    \begin{tabular}{ccc}
   \resizebox{55mm}{!}{
         	\begin{overpic}[scale=0.5]%
{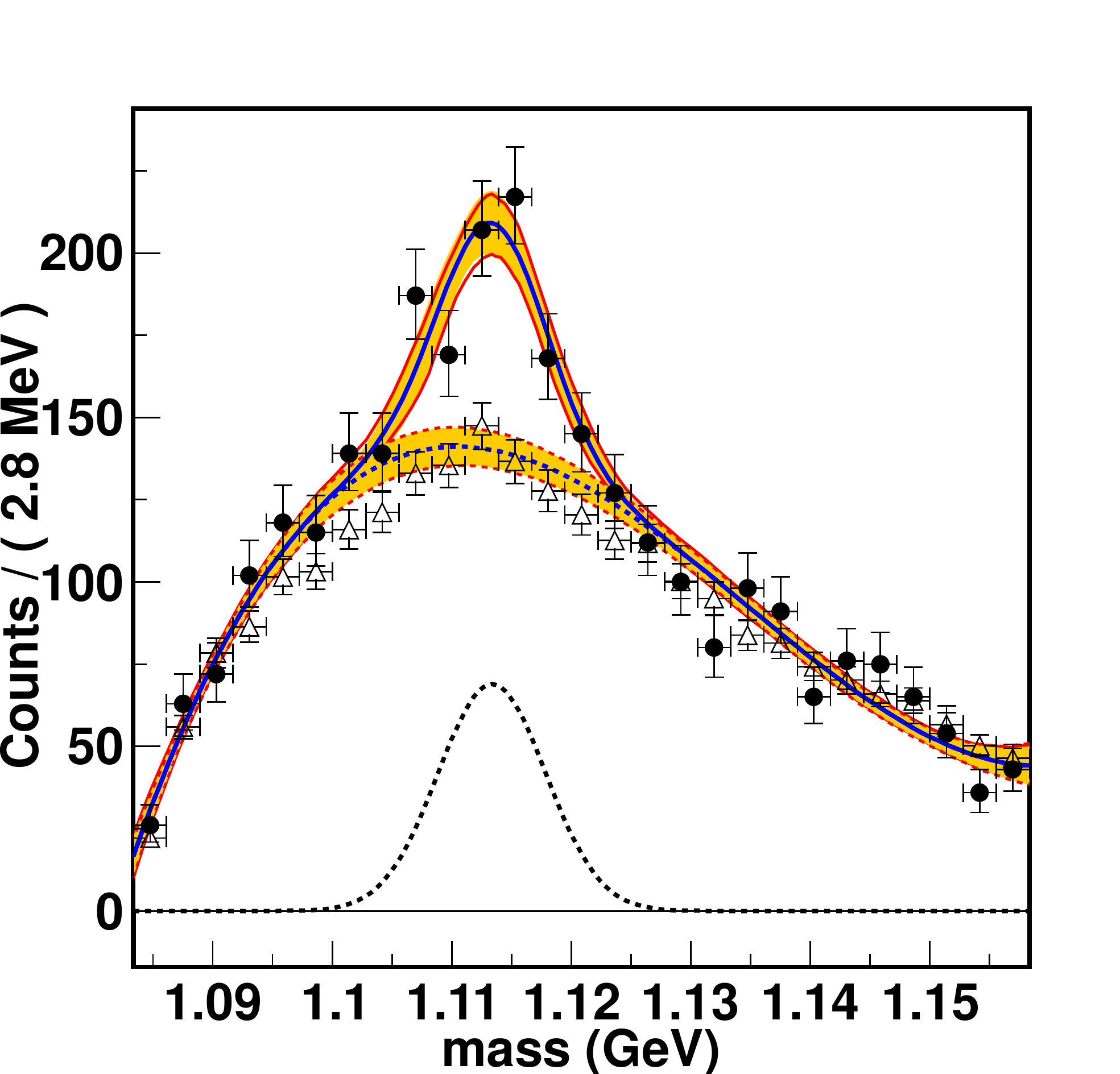}
 		  \put(170,210){\textcolor{black}{\textbf{\Large{$\chi^2/ndf=0.95$}}}}
 		  \put(220,230){\textcolor{black}{\textbf{\Large{(1a)}}}}
    \end{overpic}
} & 
   \resizebox{55mm}{!}{
            	\begin{overpic}[scale=0.5]%
{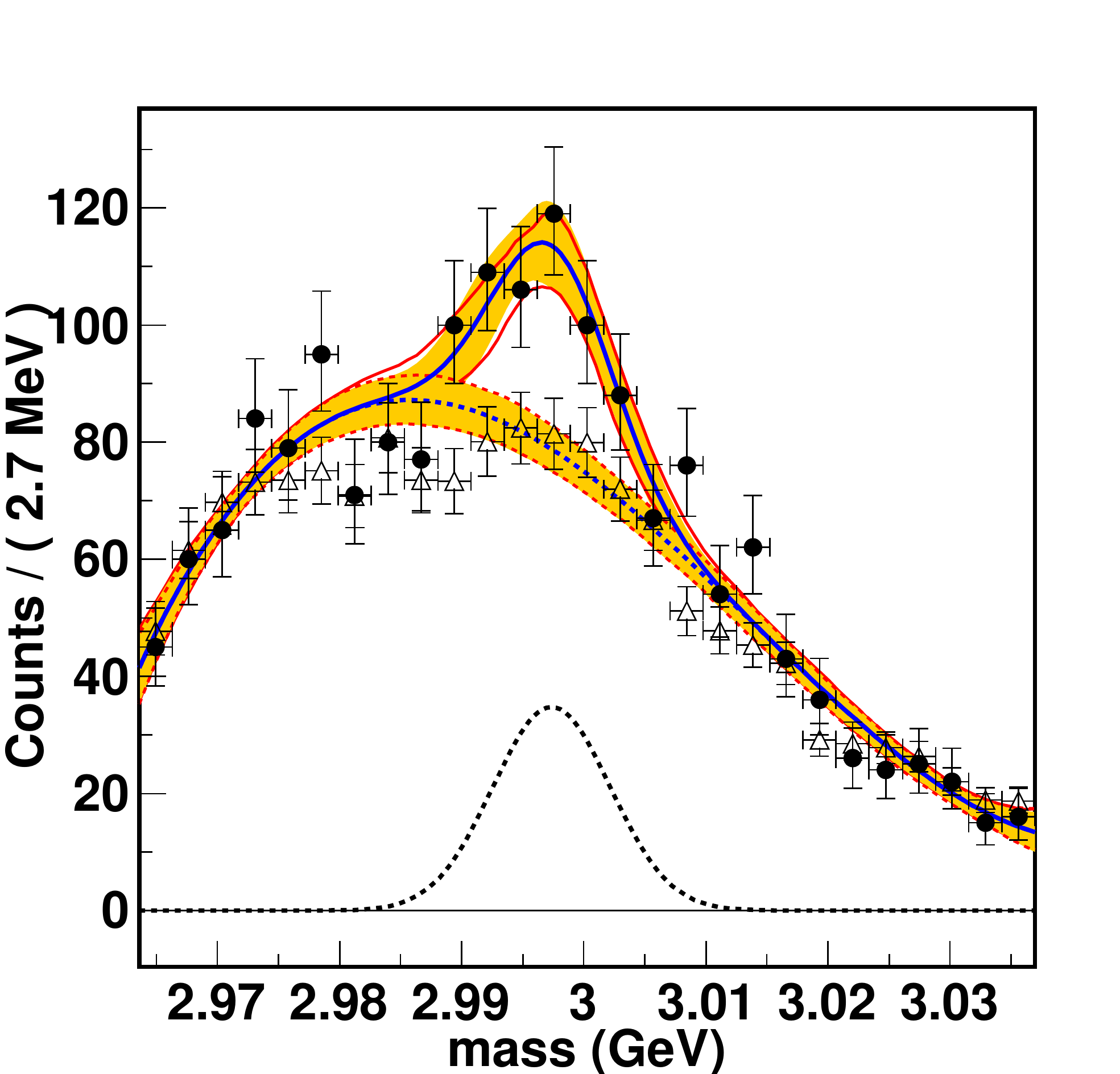}
 		  \put(170,210){\textcolor{black}{\textbf{\Large{$\chi^2/ndf=0.64$}}}}
 		  \put(220,230){\textcolor{black}{\textbf{\Large{(1b)}}}}
    \end{overpic}
} & 
   \resizebox{55mm}{!}{
            	\begin{overpic}[scale=0.5]%
{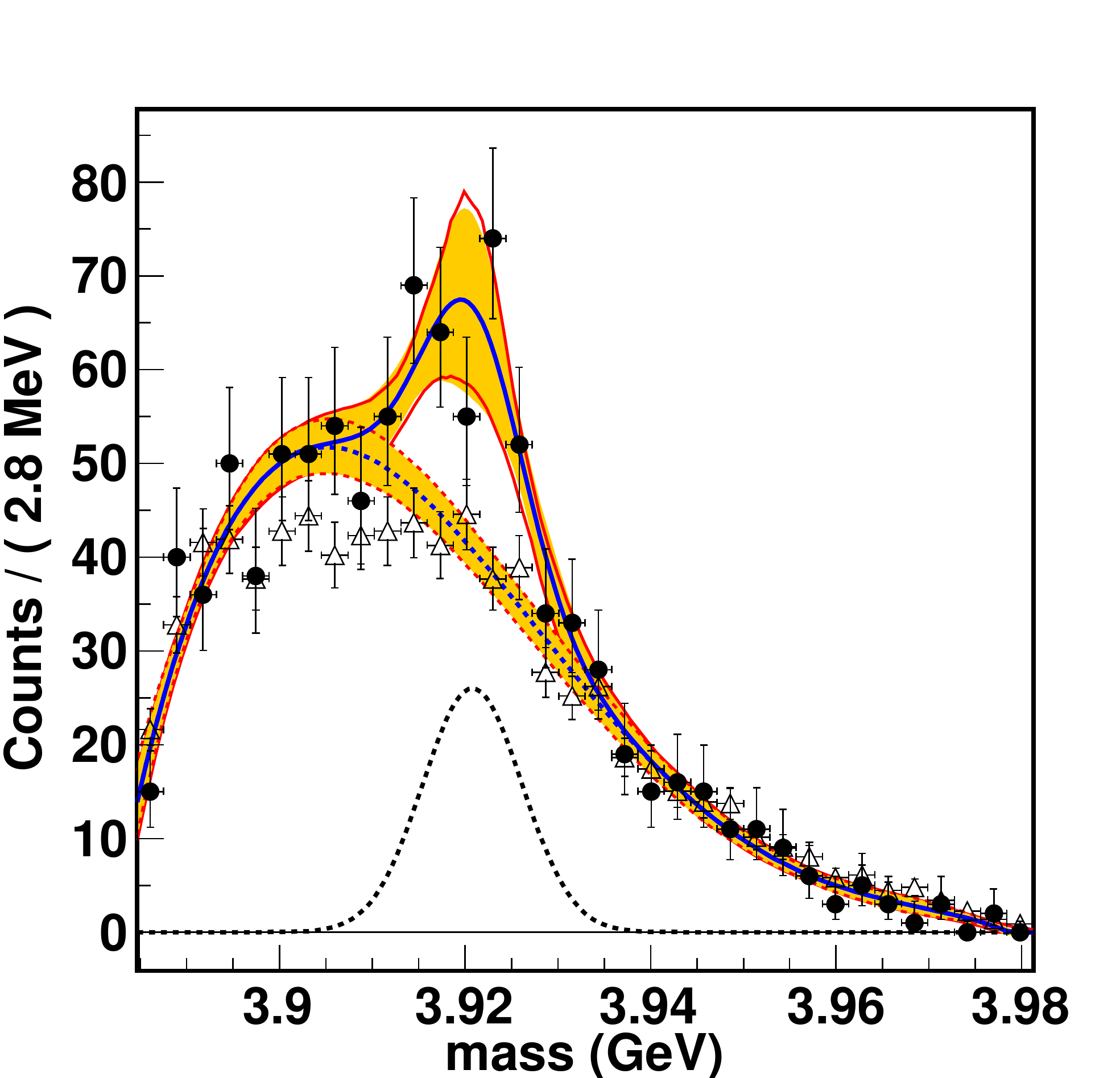}
 		  \put(170,210){\textcolor{black}{\textbf{\Large{$\chi^2/ndf=0.52$}}}}
 		  \put(220,230){\textcolor{black}{\textbf{\Large{(1c)}}}}
    \end{overpic}
} \\

   \resizebox{55mm}{!}{
         	\begin{overpic}[scale=0.5]%
{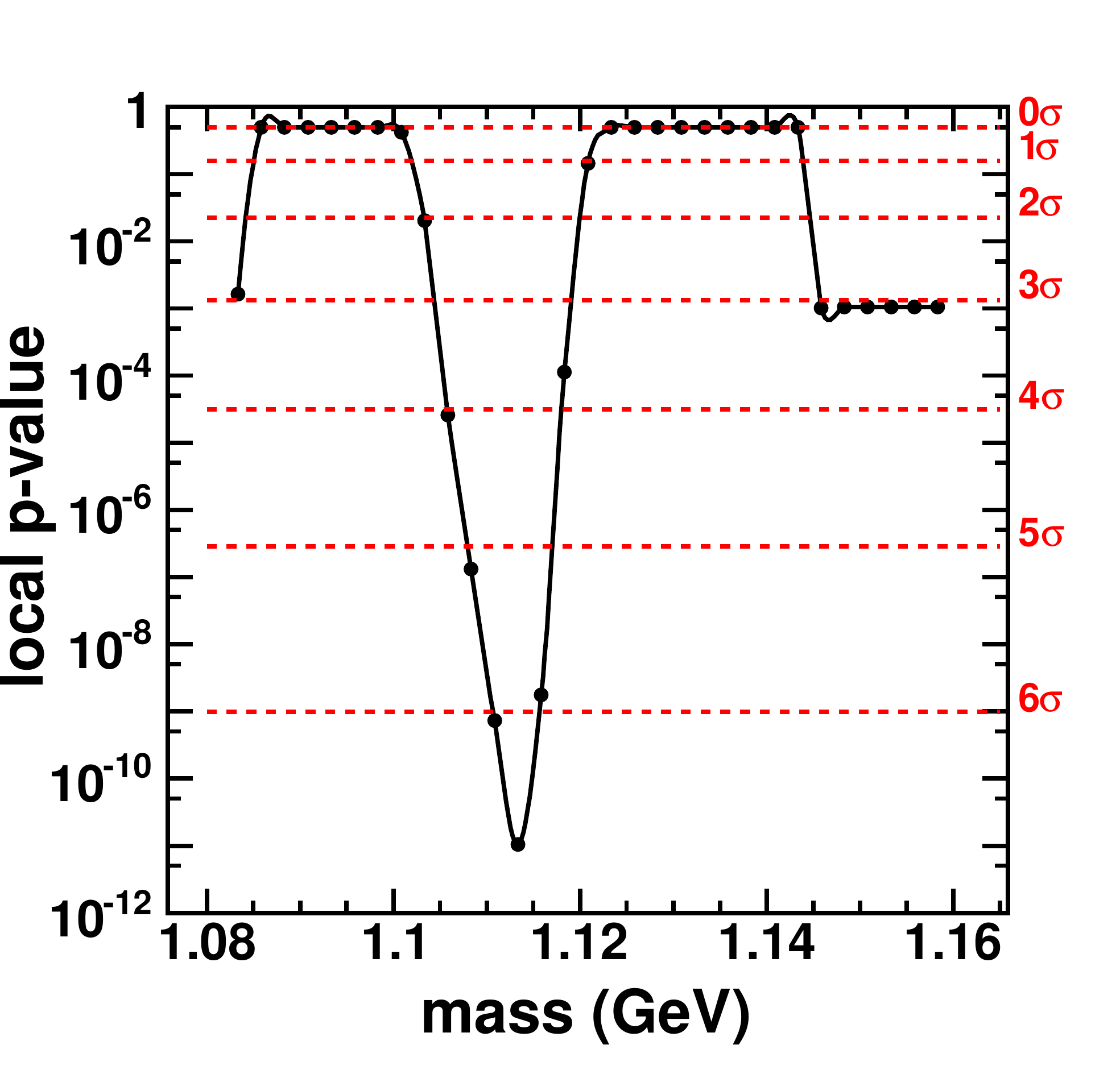}
 		  \put(220,223){\textcolor{black}{\textbf{\Large{(2a)}}}}
    \end{overpic}
} & 
   \resizebox{55mm}{!}{
            	\begin{overpic}[scale=0.5]%
{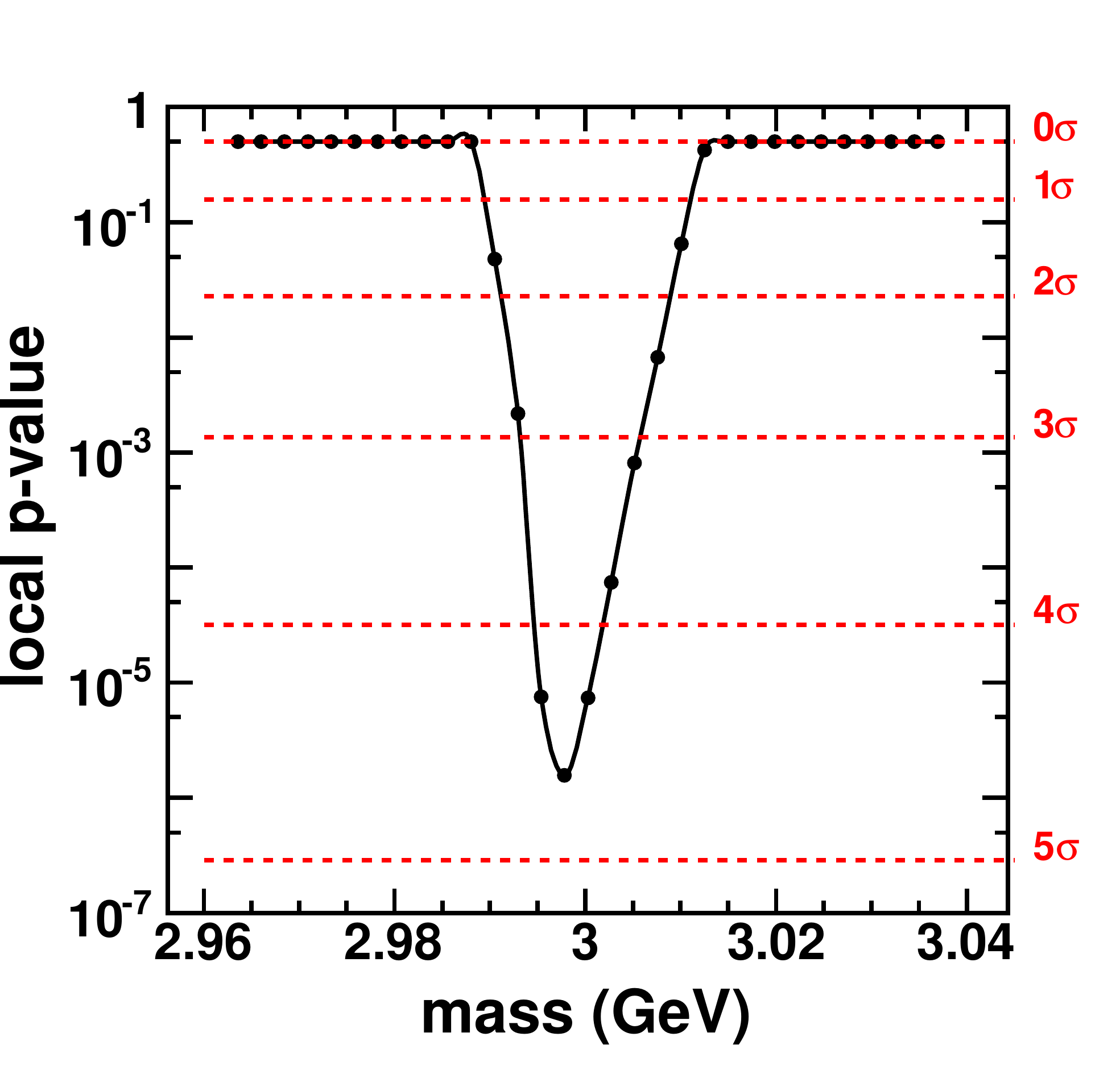}
 		  \put(220,223){\textcolor{black}{\textbf{\Large{(2b)}}}}
    \end{overpic}
} & 
   \resizebox{55mm}{!}{
            	\begin{overpic}[scale=0.5]%
{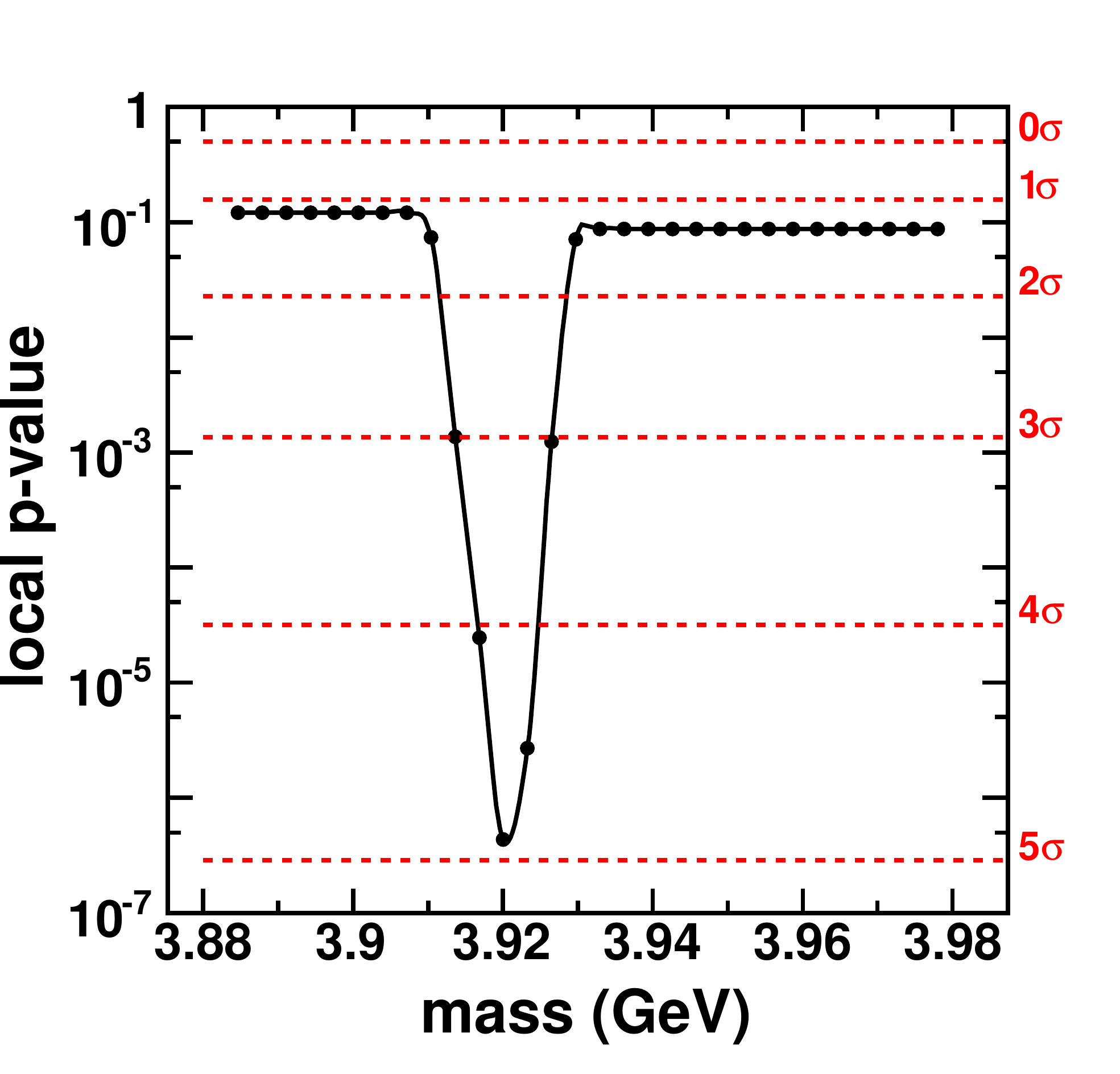}
 		  \put(220,223){\textcolor{black}{\textbf{\Large{(2c)}}}}
    \end{overpic}
} \\

    \end{tabular}

   \caption{
\label{fig:invmass}
(Color online) Invariant mass distribution for candidates of $\Lambda$, $^3_{\Lambda}$H and $^4_{\Lambda}$H, are represented by the filled circles in panels (1a), (1b), and (1c), respectively. The shaded orange region represents one standard deviation of the fitted model centered at the solid blue line. The dotted lines show the separate contributions of the signal and the background with, respectively, black and colored lines. The data represented by open triangles correspond to invariant mass distributions of the mixed event analysis. The local p-value distribution of the background-only hypothesis in full range of fit of $\Lambda$, $^3_{\Lambda}$H and $^4_{\Lambda}$H, are shown in panels (2a), (2b) and (2c), respectively. The red dashed lines illustrate the p-values corresponding to significances of 1$\sigma$, 2$\sigma$, 3$\sigma$, 4$\sigma$, 5$\sigma$ and 6$\sigma$.} %

\end{figure*}
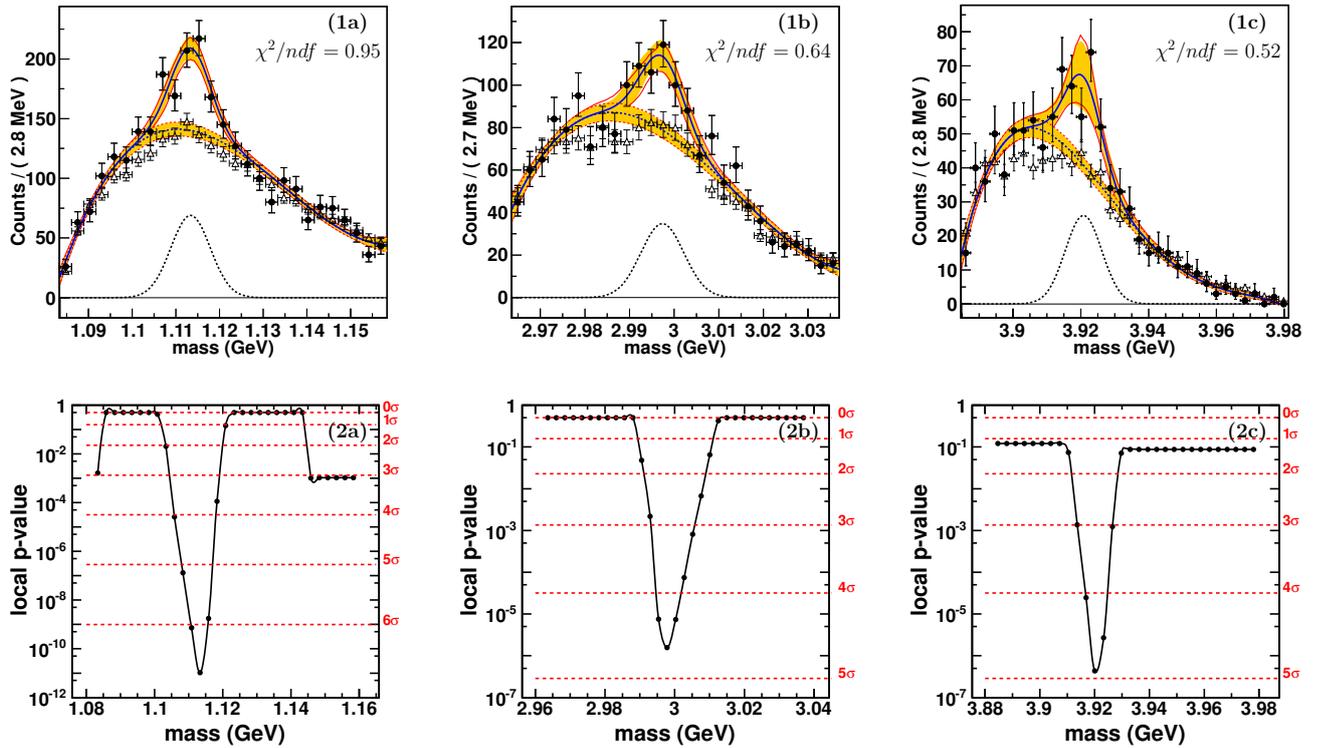

The top panels of Figure \ref{fig:invmass} show the invariant mass distributions for  p~+~$\pi^-$, $^3$He~+~$\pi^-$, and $^4$He~+~$\pi^-$ candidates. In each distribution, a peak at around the mass of $\Lambda$, $^3_{\Lambda}$H and $^4_{\Lambda}$H can be seen. 
The background contributions to the invariant mass distributions initially were estimated with the event mixing method \cite{Drijard:1984pe,Crochet:2001qd}. The distribution from the mixing event method is obtained by combining uncorrelated candidate daughter particles for the mother state of interest that belong to distinct events. The obtained mixed event distributions are then normalized for purpose of comparison with the invariant mass of the mother state of interest. The scaling factors are calculated as the ratio of the total integral of the two distributions in a mass region in which the signal contribution is considered to be negligible. The background contribution estimated via mixed event analysis is shown in Figure \ref{fig:invmass} by the open triangles.
 
The separated signal and background contributions have been estimated via the binned maximum likelihood fitting method. A signal in the invariant mass distributions is represented by a Gaussian probability density function.  The background contribution was then modeled by a Chebychev polynomial probability density function of the first kind. An extended maximum likelihood estimator allowed us to obtain the share of signal and background in each invariant mass spectrum. 

\begin{figure*}[thb]
\centering
    \begin{tabular}{ccc}
   \resizebox{55mm}{!}{
         	\begin{overpic}[scale=0.5]%
{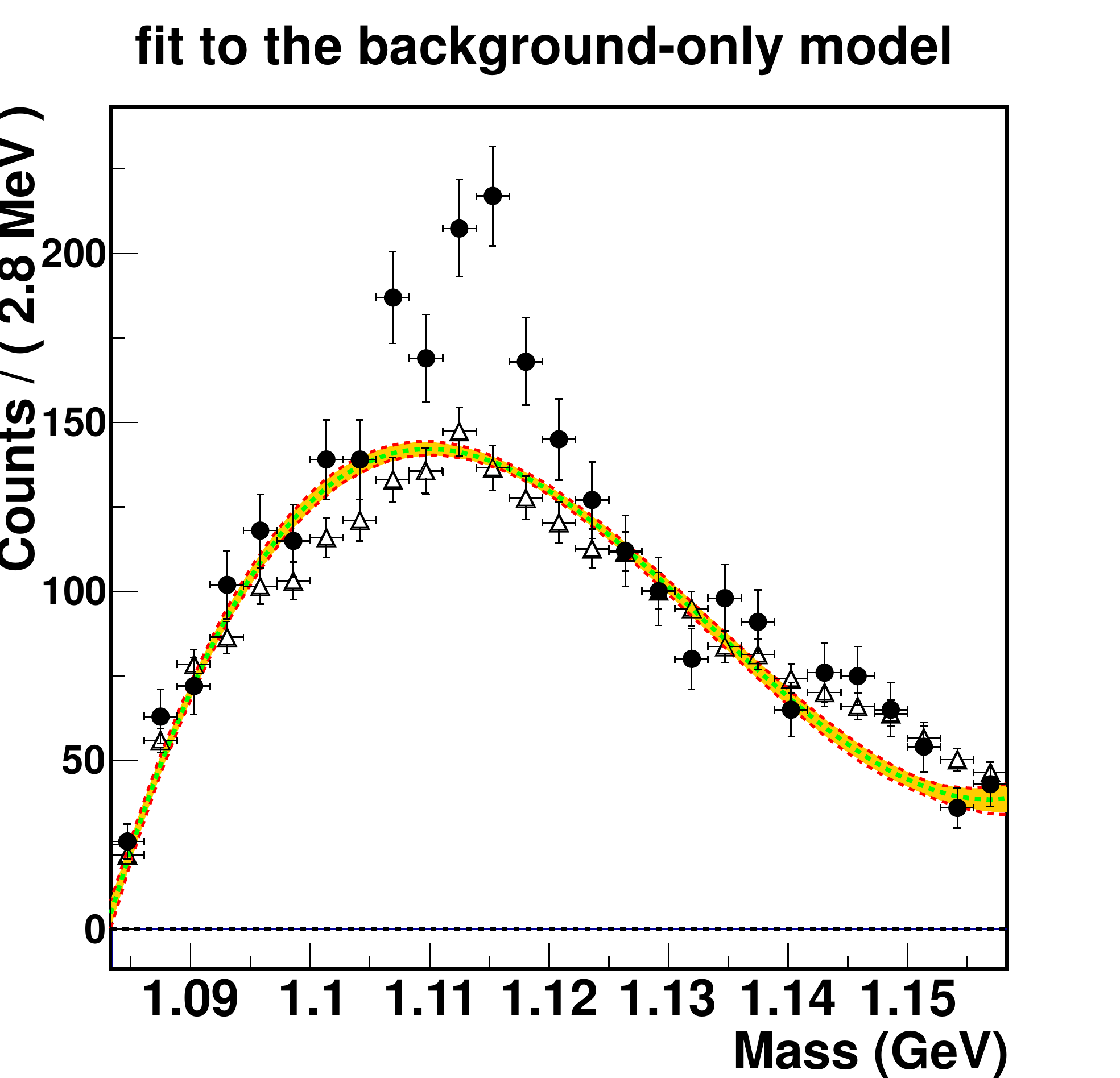}
 		  \put(220,230){\textcolor{black}{\textbf{\Large{(a)}}}}
    \end{overpic}
} & 
   \resizebox{55mm}{!}{
            	\begin{overpic}[scale=0.5]%
{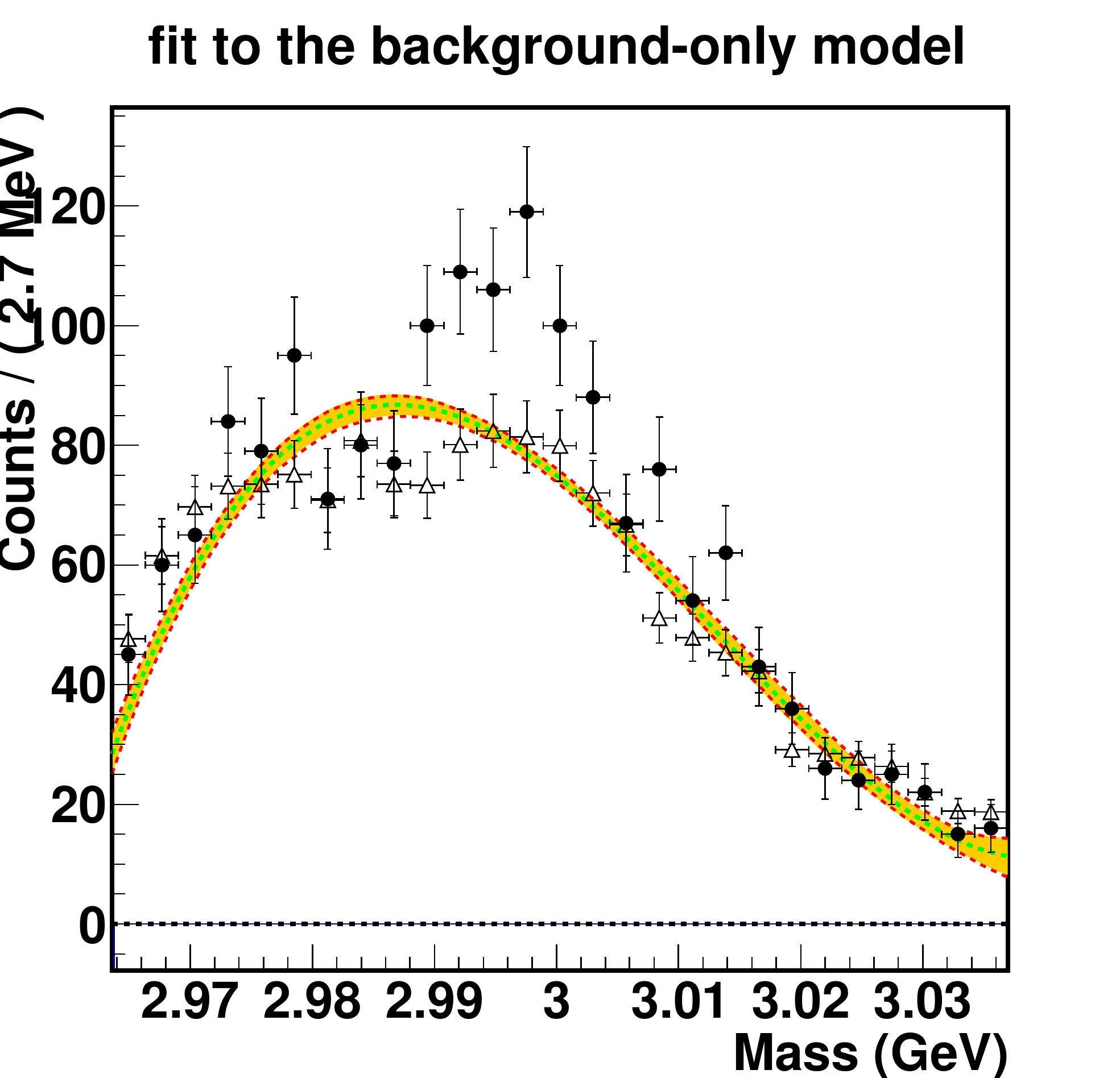}
 		  \put(220,230){\textcolor{black}{\textbf{\Large{(b)}}}}
    \end{overpic}
} & 
   \resizebox{55mm}{!}{
            	\begin{overpic}[scale=0.5]%
{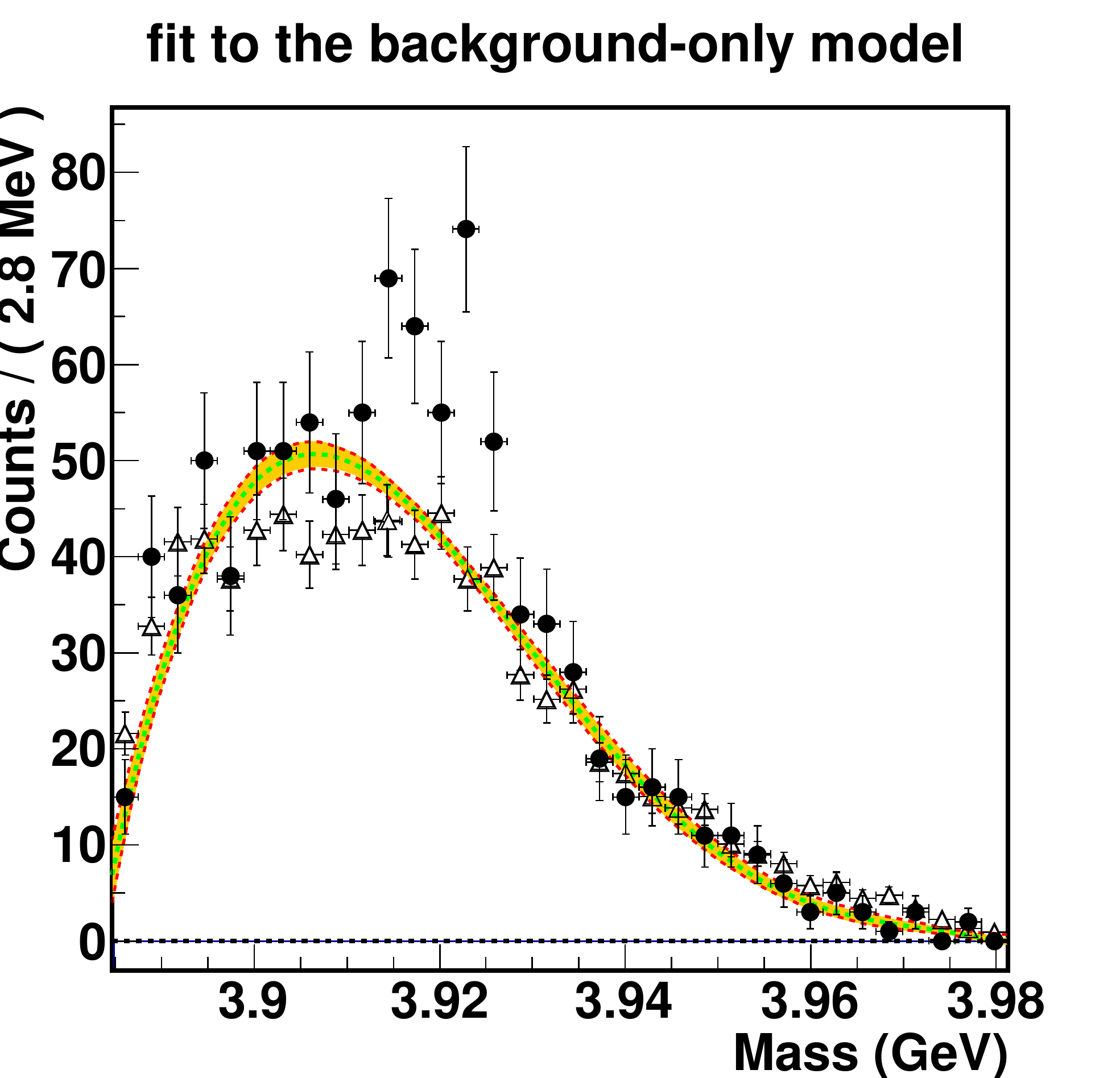}
 		  \put(220,230){\textcolor{black}{\textbf{\Large{(c)}}}}
    \end{overpic}
} \\

    \end{tabular}

   \caption{
\label{fig:invmass_backgroundonly}
(Color online) Fitting of background-only hypothesis of the mixed event invariant mass distribution of the candidates of $\Lambda$, $^3_{\Lambda}$H and $^4_{\Lambda}$H, is shown in panels (a), (b), and (c). The shaded orange region represents one standard deviation of the fitted model centered at the solid green line. The data represented by open triangles corresponds to invariant mass distributions of the mixed event analysis. In addition, the normal invariant mass distribution of the mother state of interest is retained in order to demonstrate the divergence of the fit based on the background-only hypothesis.}
         
\end{figure*}

Resultant values of the signal and background contributions are listed in Table \ref{tab:mass_sig}. In addition, a hypothesis test was applied to determine the significance of the signal, in which H0 corresponds to the background-only hypothesis while H1 does so for the signal-plus-background. A profiled maximum likelihood ratio test gives us a significance of 6.7$\sigma$, 4.7$\sigma$, 4.9$\sigma$ for $\Lambda$ hyperon, $_\Lambda^3$H and $_\Lambda^4$H, respectively. The procedures for fitting and hypothesis testing are based on the RooFit and RooStats frameworks \cite{cite:roofit,cite:roostats}. Also shown in the bottom panels of Figure \ref{fig:invmass} is the local p-value of the background-only hypothesis as a function of the mass value of the possible signal of the mother state of interest. For each data point, the local p-value is calculated by the profiled maximum likelihood ratio test in which the mean mass value of the signal contribution is fixed in the likelihood function of the H1 hypothesis. The distribution is then obtained by scanning the mass region. The maximum of signal significance can be deduced and the presence of the signal contribution into the invariant mass distribution of $\Lambda$ hyperon, $_\Lambda^3$H and $_\Lambda^4$H demonstrated. Figure \ref{fig:invmass_backgroundonly} also shows the fit of the background-only hypothesis H0 with the invariant mass distribution of the mixed event analysis obtained from the experimental data.

\begin{table*}[tHb] %
\centering
\caption{\label{tab:mass_sig} Summary of the results obtained by fitting the invariant mass distributions. The parameter $N_{sig}$ represents the integral of the signal contribution, while  $N_{bg}$ that of the background contributions. The mean value and $\sigma $ of the Gaussian model are referred to as $\bar{m}$  and $\sigma_{m}$, respectively. Parameters, $a_0$, $a_1$, $a_2$ and $a_3$, are the coefficients of the Chebychev polynomial. 
 }
\begin{tabular}{cccc}
\toprule
Fitted values & $\Lambda$  & $^3_{\Lambda}$H & $^4_{\Lambda}$H \\
\midrule
$N_{sig}$ & 280 $\pm$ 63 & 154 $\pm $ 49 & 123 $\pm$ 33 \\
$\bar{m}$ (MeV/c$^2$) & 1113.3 $\pm$ 0.8 & 2997.4 $\pm$ 1.2 & 3920.8 $\pm$ 1.2 \\
$\sigma_{m}$ (MeV/c$^2$) & 4.5 $\pm$ 0.9 & 4.8 $\pm$ 1.3 & 5.4 $\pm$ 1.2 \\
$N_{bg}$ & 2609 $\pm$ 79 & 1590 $\pm $ 62 & 841 $\pm$ 43 \\
 $a_0$ & -0.106$\pm $0.034 & -0.487$\pm $0.042 & -0.947$\pm $0.034 \\
 $a_1$ & -0.621$\pm $0.044 & -0.462$\pm $0.062 & -0.467$\pm $0.065 \\
 $a_2$ & 0.279$\pm $0.040 & 0.212$\pm $0.047 & -0.590$\pm $0.068 \\
 $a_3$ &  &  & -0.232$\pm $0.057 \\
\bottomrule

\end{tabular}
\end{table*}

\section{Analysis: Lifetime estimation}

After reconstructing the invariant mass, the lifetime of $\Lambda $, $^3_\Lambda $H and  $^4_\Lambda $H was extracted. The signal contribution was determined by subtracting the background contribution from the signal-plus-background contribution. For this purpose, two data sets are built. The signal-plus-background contribution was deduced in the fitted peak region within the fitted values of $\bar{m} \pm 2\sigma_m$, and the background-only data set was extracted from the adjacent sideband region within the intervals of [$\bar{m} -4\sigma_m,\bar{m} -2\sigma_m$] and [$\bar{m} +2\sigma_m,\bar{m} +4\sigma_m$]. Thanks to the mixed event analysis, the portion of background inside the signal-plus-background region can be estimated deductively and used to normalize the integrated area of the sideband region. Each of the sideband data sets is then associated with a normalization factor $N_{bl}$ and $N_{bh}$.

\begin{figure*}[htb]
\centering
   \begin{tabular}{ccc}

   \resizebox{55mm}{!}{
         	\begin{overpic}[scale=0.5]%
{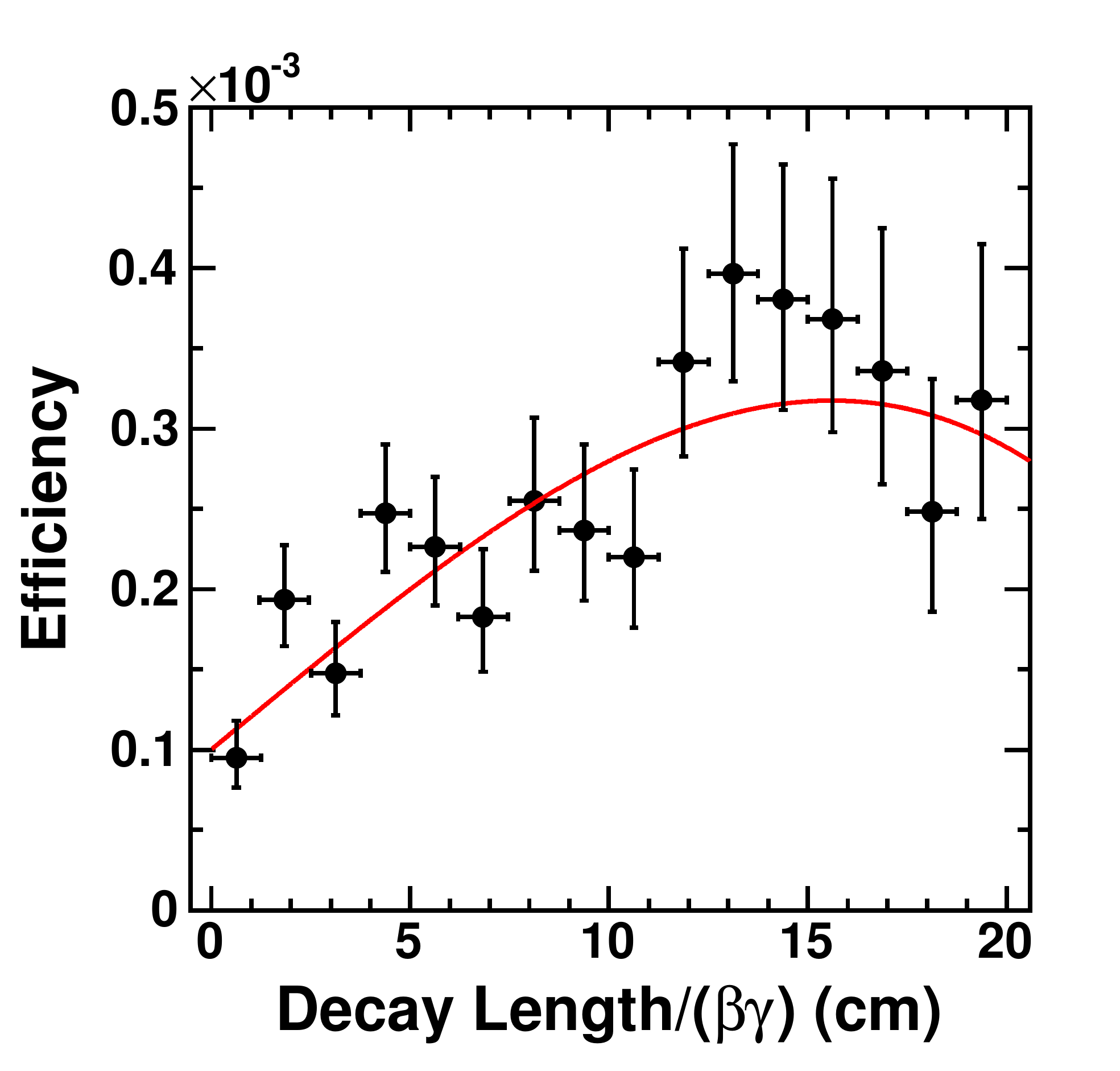}
 		  \put(230,230){\textcolor{black}{\textbf{\Large{(a)}}}}
    \end{overpic}
} & 
   \resizebox{55mm}{!}{
            	\begin{overpic}[scale=0.5]%
{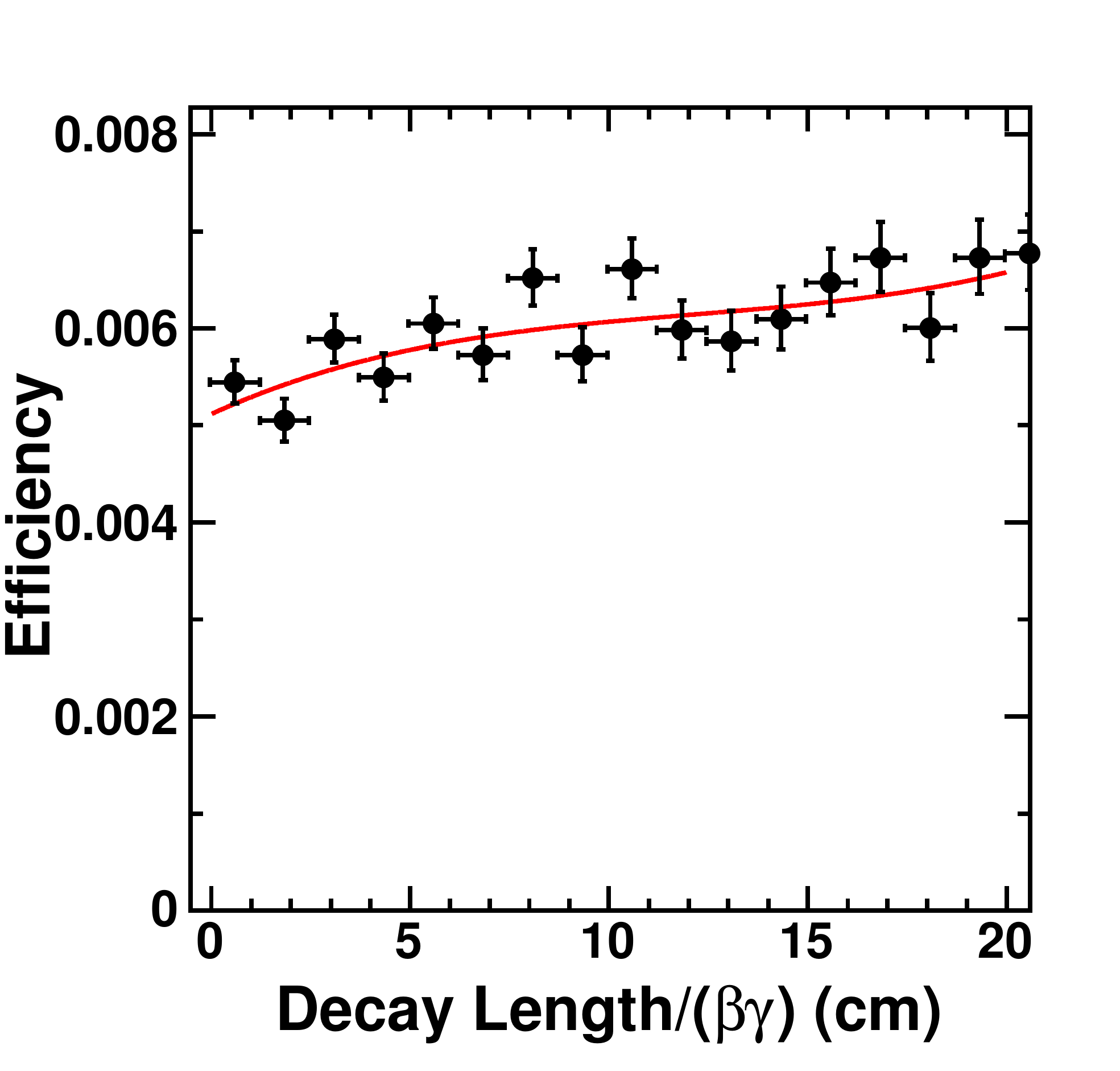}
 		  \put(230,230){\textcolor{black}{\textbf{\Large{(b)}}}}
    \end{overpic}
} & 
   \resizebox{55mm}{!}{
            	\begin{overpic}[scale=0.5]%
{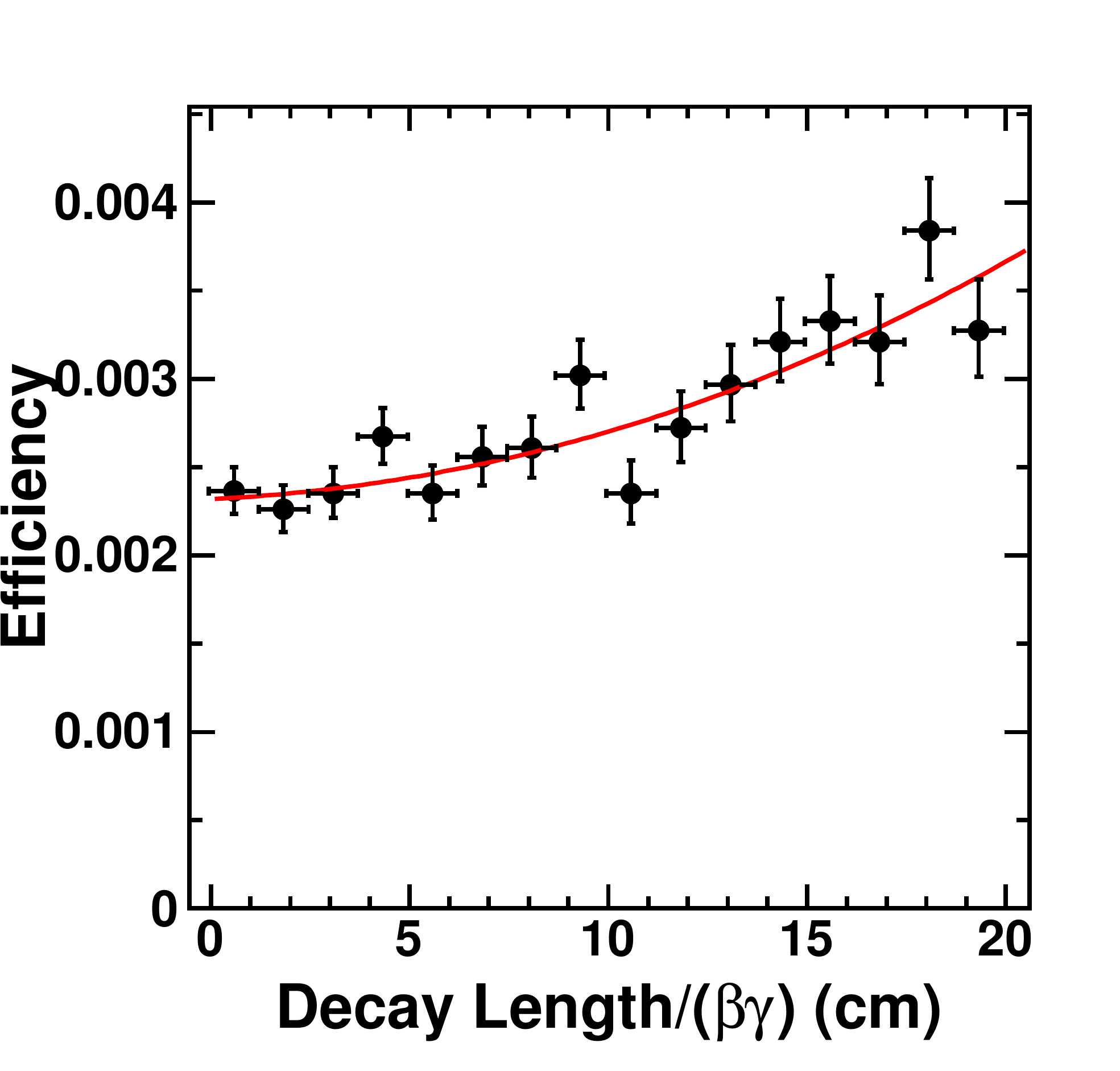}
 		  \put(230,230){\textcolor{black}{\textbf{\Large{(c)}}}}
    \end{overpic}
} \\

    \end{tabular}

  \caption{\label{fig:eff_corr} Acceptance and reconstruction efficiency correction function obtained via Monte Carlo simulations for $\Lambda$, $^3_{\Lambda}$H and $^4_{\Lambda}$H, are shown in panels (a), (b), and (c) respectively. A polynomial function is used to model the efficiency correction (shown as a red solid line).}
         
\end{figure*}

In addition, the longitudinal vertex position had to be situated more than 6 cm distant from the target in order to avoid any contamination from reaction events at the target and at the tracking detector immediately behind the target. In each data set, the proper decay time $t=l/(\beta\gamma\,c)$ is then calculated in the rest frame of the mother state of interest from the measured decay length $l$, where $\beta\gamma\,c=p/m$, $p$ and $m$ the momentum and the mass of the mother state of interest. The yield at different decay lengths was also corrected by the deduced acceptance and reconstruction efficiency based on full Monte Carlo simulations, shown in Figure \ref{fig:eff_corr}. A polynomial function was utilized to model the efficiency corrections applied to the data sets.

\begin{figure*}[htb]
\centering
    \begin{tabular}{ccc}
   \resizebox{55mm}{!}{
      	\begin{overpic}[scale=0.5]%
 {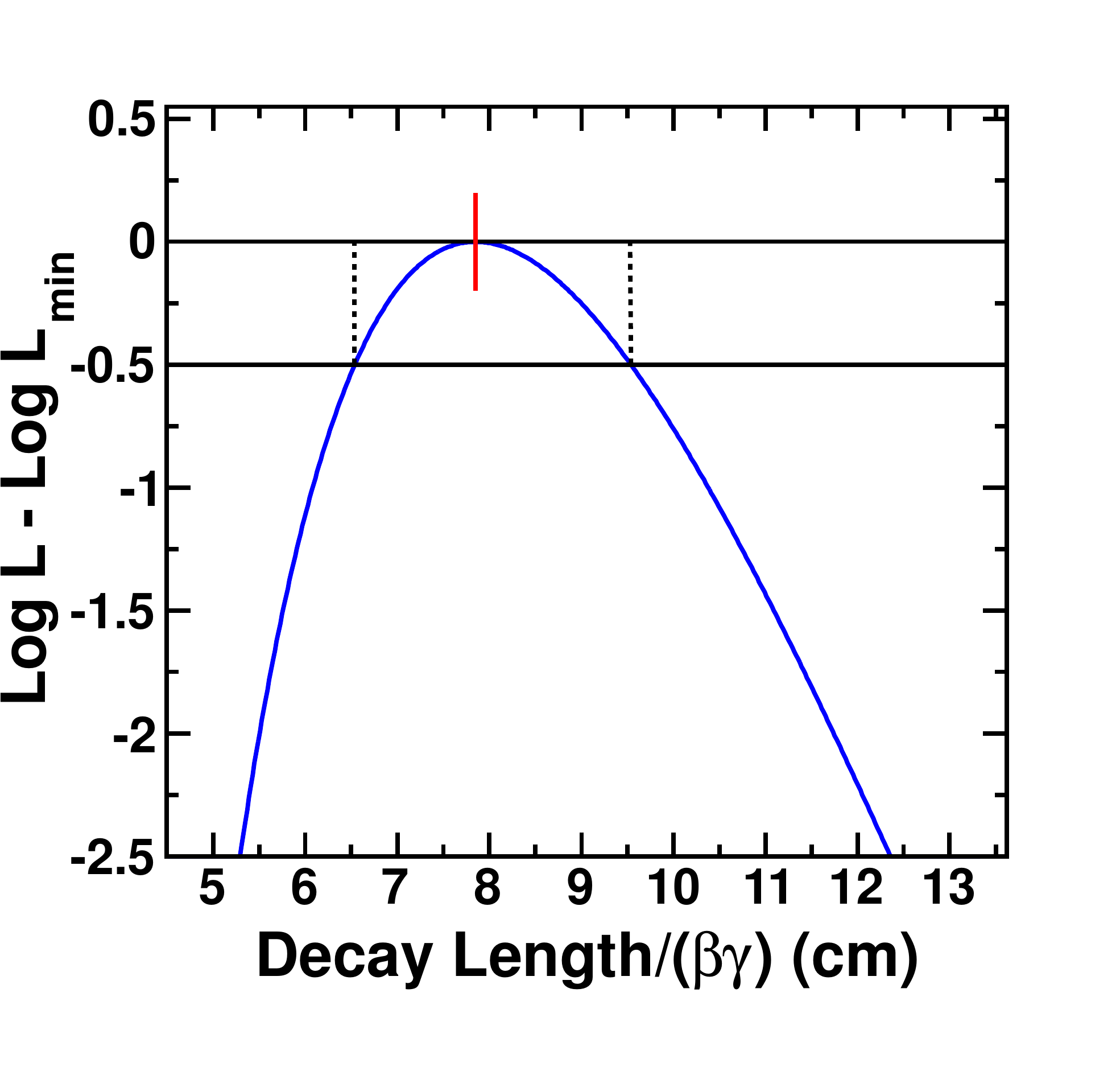}	
       		  \put(120,220){\textcolor{black}{\textbf{\large{7.850 cm}}}}
 		  \put(220,225){\textcolor{black}{\textbf{\Large{(a1)}}}}
        \end{overpic}	
} &

   \resizebox{55mm}{!}{
         	\begin{overpic}[scale=0.5]%
 {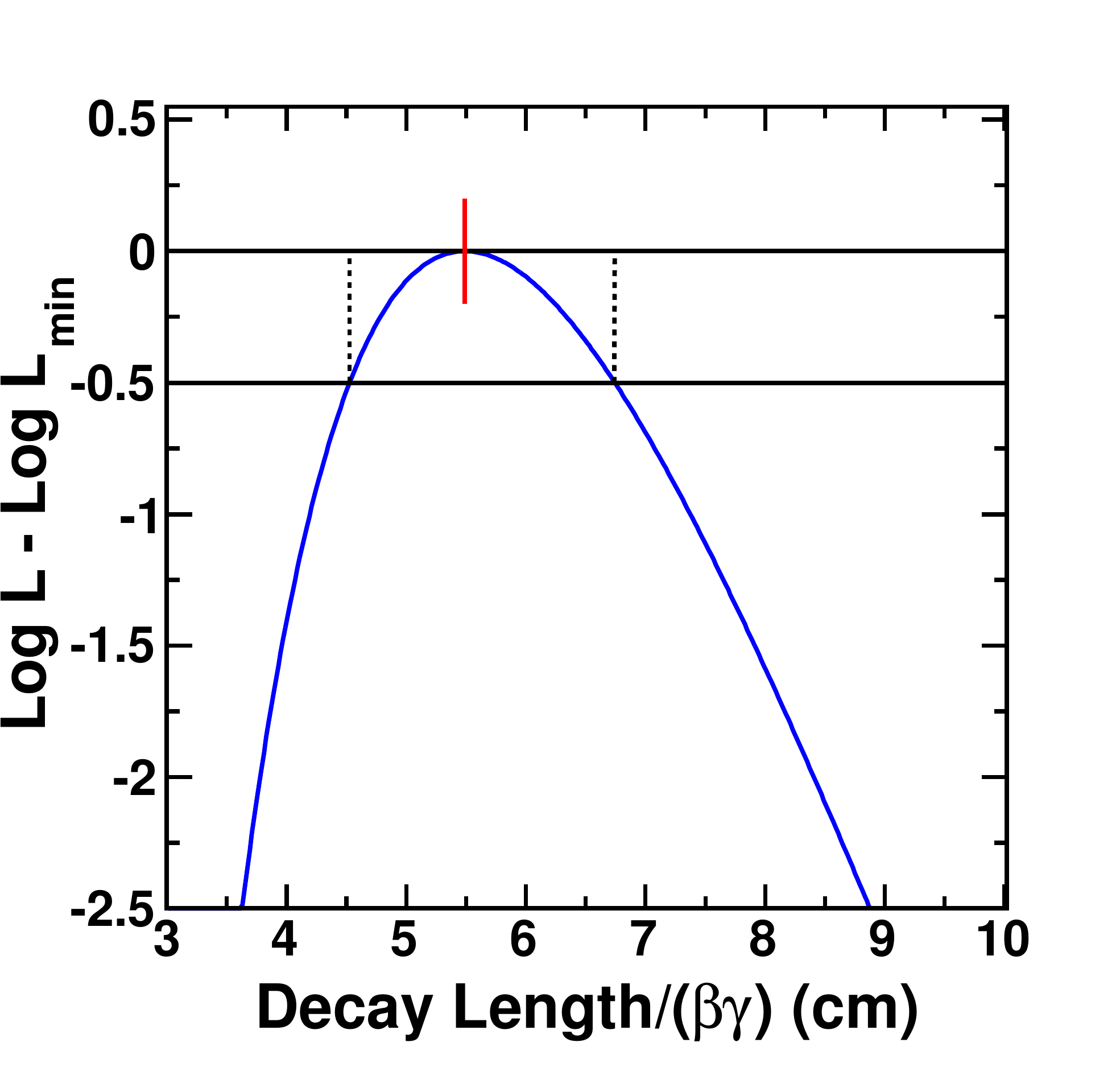}
           		  \put(118,219){\textcolor{black}{\textbf{\large{5.488 cm}}}}
  		  \put(220,225){\textcolor{black}{\textbf{\Large{(b1)}}}}

         \end{overpic}	
} & 

   \resizebox{55mm}{!}{      	
   \begin{overpic}[scale=0.5]%
{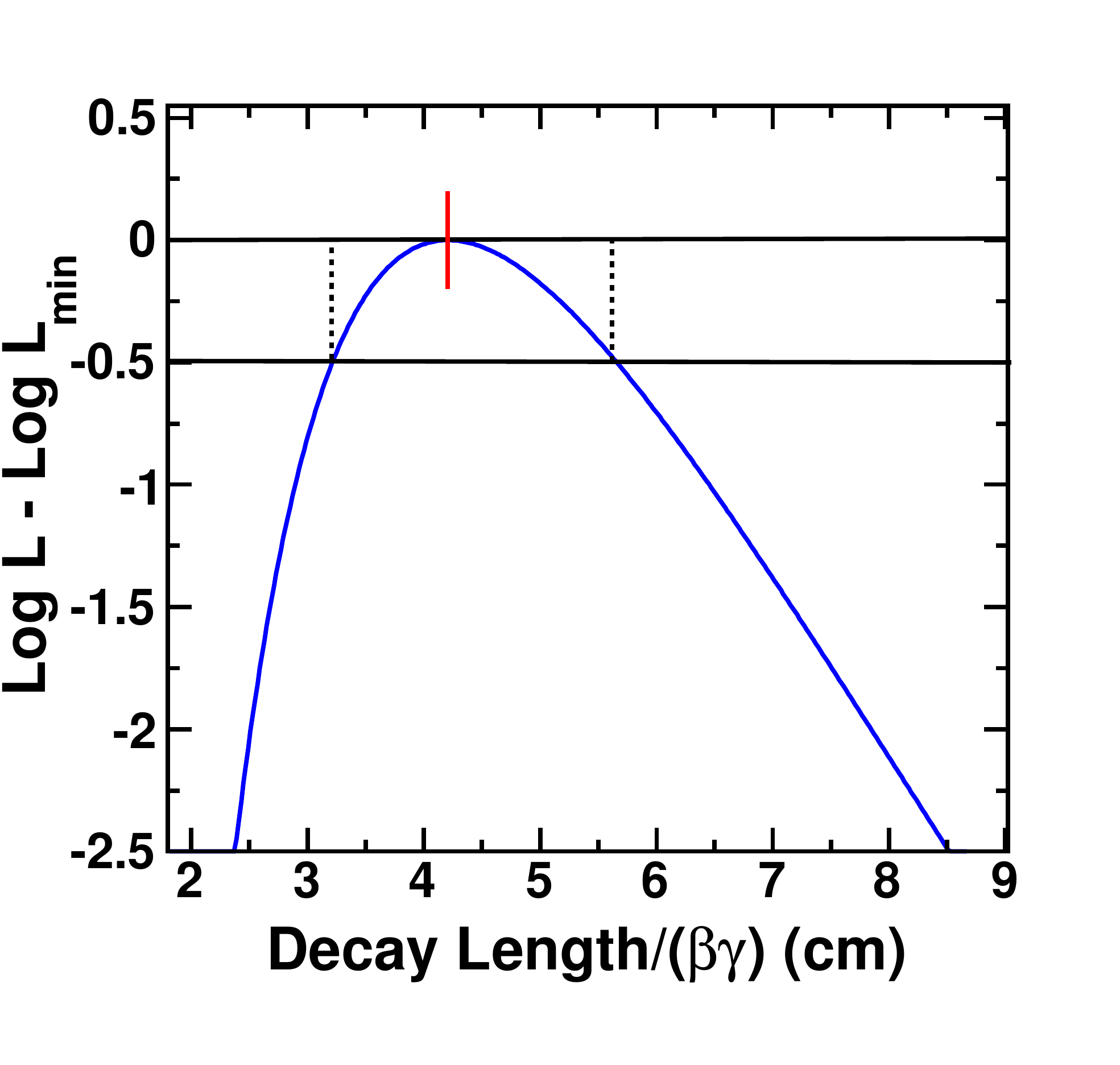}
          		  \put(123,220){\textcolor{black}{\textbf{\large{4.208 cm}}}}
 		  \put(220,225){\textcolor{black}{\textbf{\Large{(c1)}}}}
        \end{overpic}	
} \\

   \resizebox{55mm}{!}{
      	\begin{overpic}[scale=0.5]%
            {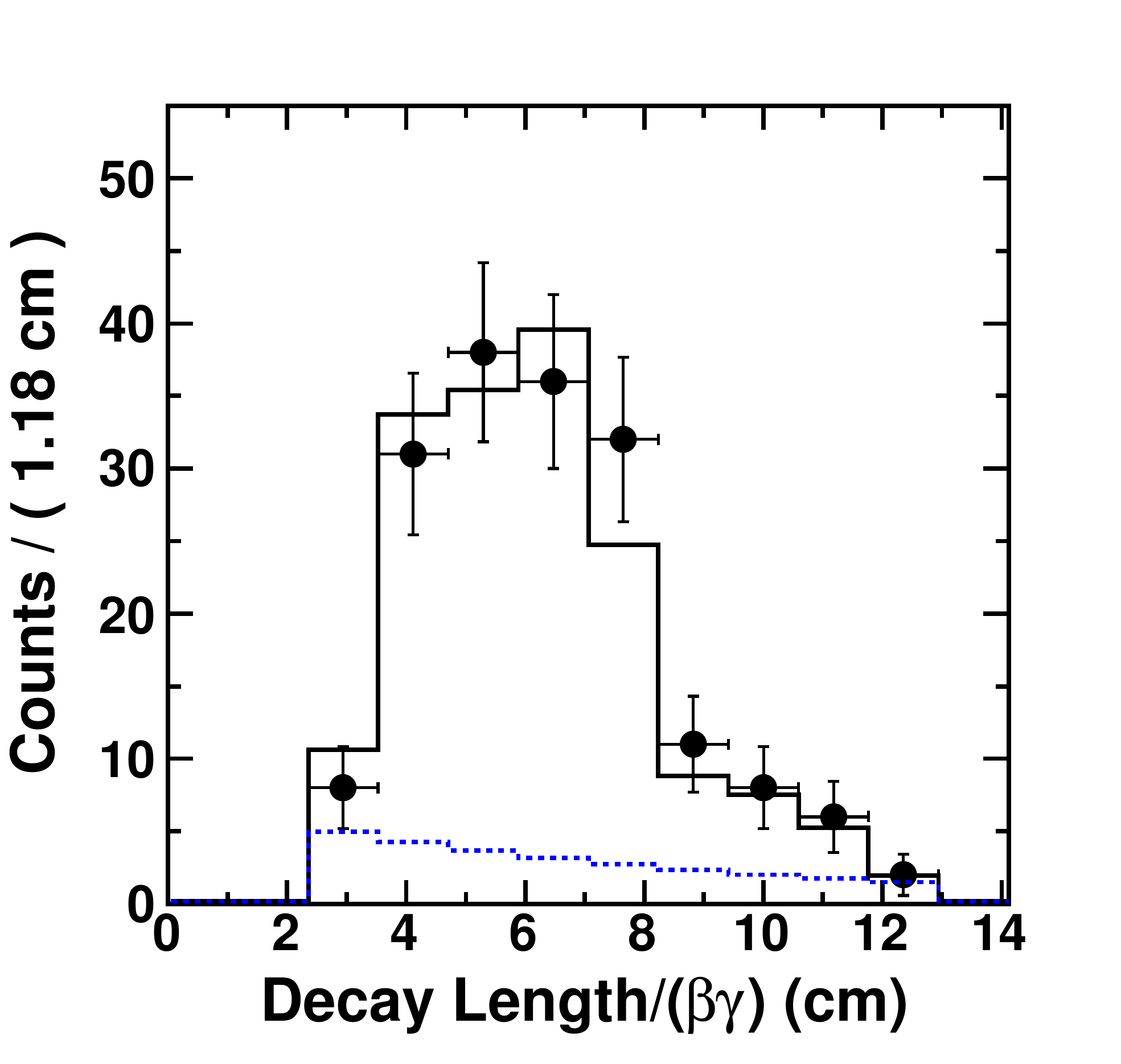}
       		  \put(60,225){\textcolor{black}{\textbf{\large{GoF : $\chi^2/ndf=0.43$}}}}
       		  \put(60,205){\textcolor{black}{\textbf{\large{p-value : Prob = 0.922}}}}
 		  \put(220,225){\textcolor{black}{\textbf{\Large{(a2)}}}}
        \end{overpic}	
    } &

   \resizebox{55mm}{!}{
         	\begin{overpic}[scale=0.5]%
{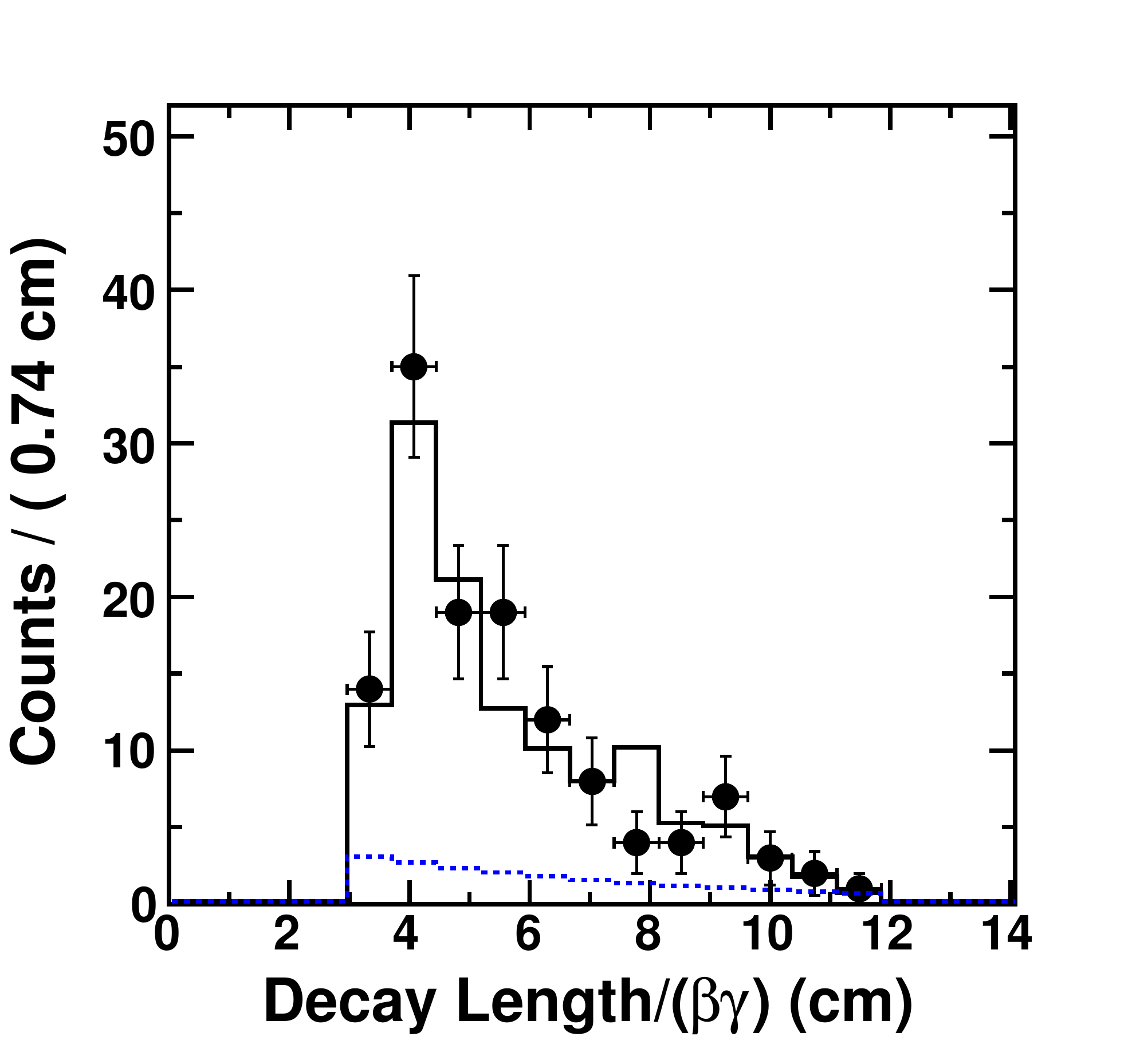}
            		  \put(60,225){\textcolor{black}{\textbf{\large{GoF : $\chi^2/ndf=1.14$}}}}
            		  \put(60,205){\textcolor{black}{\textbf{\large{p-value : Prob = 0.320}}}}
    		  \put(220,225){\textcolor{black}{\textbf{\Large{(b2)}}}}
      \end{overpic}	
    } &

   \resizebox{55mm}{!}{
         	\begin{overpic}[scale=0.5]%
   {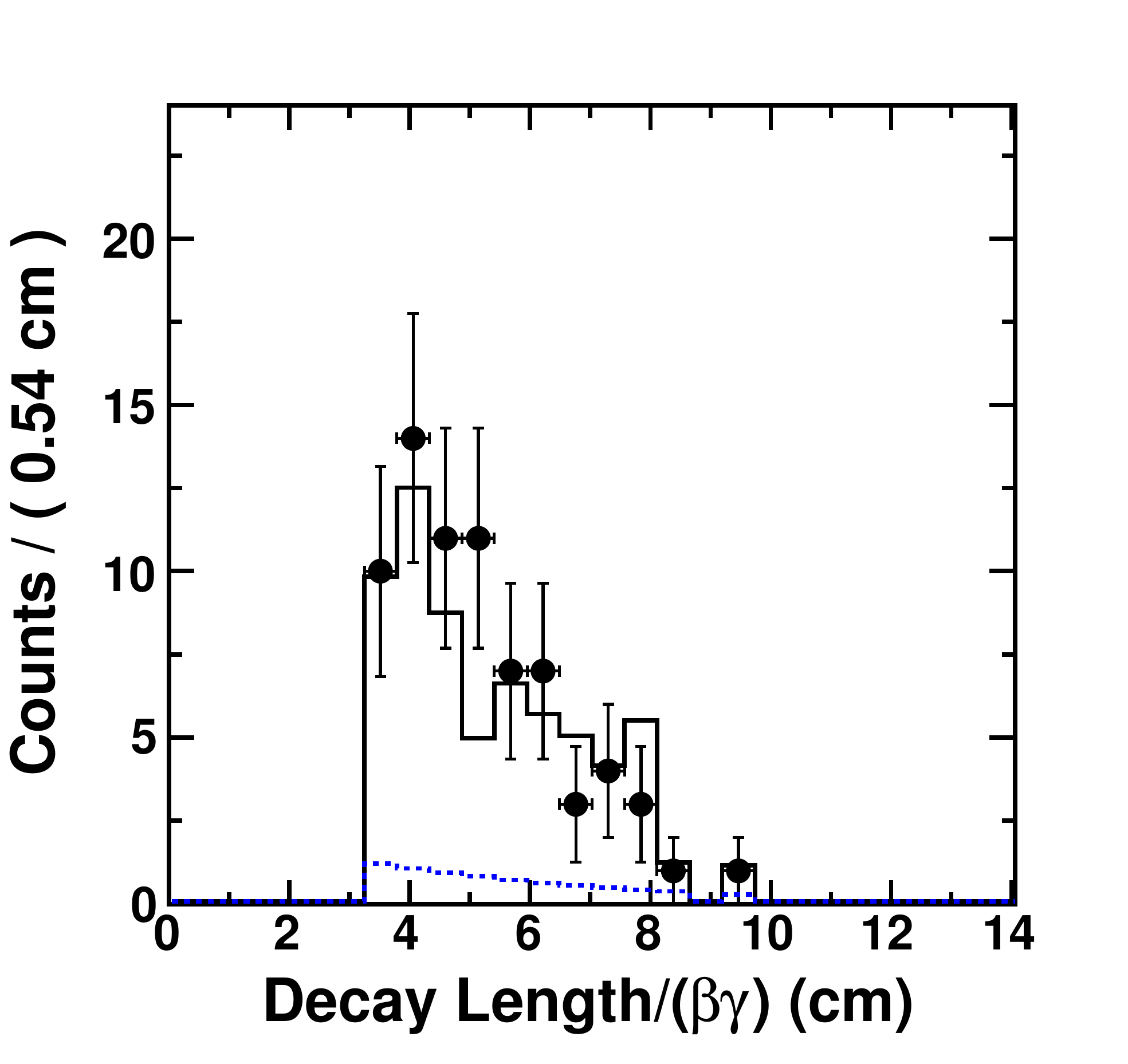}
           		  \put(60,225){\textcolor{black}{\textbf{\large{GoF : $\chi^2/ndf=0.71$}}}}
           		  \put(60,205){\textcolor{black}{\textbf{\large{p-value : Prob = 0.732}}}}
 		  \put(220,225){\textcolor{black}{\textbf{\Large{(c2)}}}}
           \end{overpic}	
      }
      \\
      
    \end{tabular}

   \caption{ 
   (Color online) Profiled likelihood ratio for interval estimation of $\Lambda$ hyperon (a1), $^3_{\Lambda}$H (b1), and $^4_{\Lambda}$H hypernuclei (c1). Interval estimation for 1 standard deviation is shown on each profiled likelihood ratio. {\it Binned} decay length distributions of the signal region of $\Lambda$ hyperon (a2), $^3_{\Lambda}$H (b2), and $^4_{\Lambda}$H hypernuclei (c2) with the fitted model, which included the exponential function resulted from the {\it unbinned} maximum likelihood fit and the background contribution estimated by the sidebands. The black line represents the fitted model, while the blue dotted line represents the contribution of the exponential function.
   }
    \label{fig:profile}     
\end{figure*}
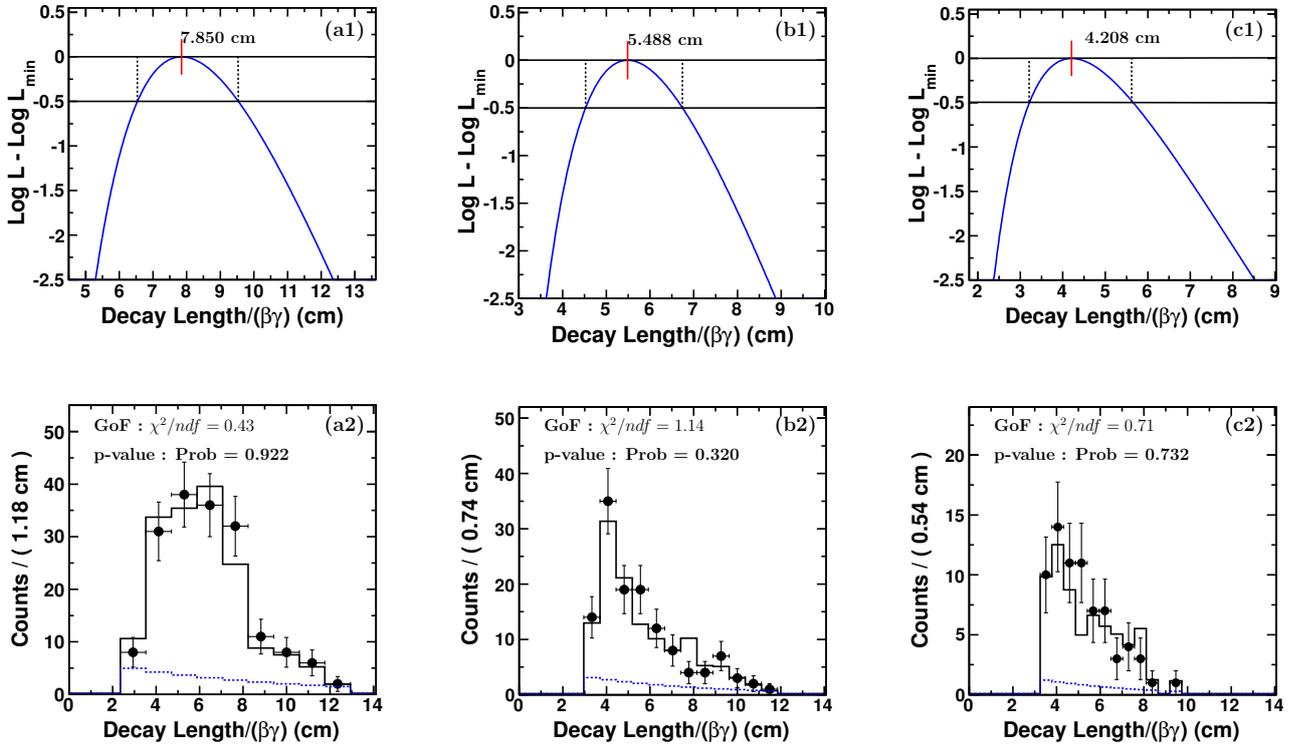

The lifetime values were then extracted utilizing an {\it unbinned} maximum likelihood fitting method. The {\it unbinned} analysis is chosen because of statistical limitations and in order to overcome the loss of information and arbitrariness of the binning. The distribution of the proper decay time $l/(\beta\gamma\,c)$ with a lifetime $\tau$ is modeled by an exponential probability density function, which gives a log-likelihood function as

\begin{equation}
  \mathcal{L} = \displaystyle \prod_i\ [\, \frac{1}{c\,\tau} e^{-l_i/(\beta_i\gamma_i\,c\,\tau)}\, ]^{w_i'} 
\end{equation}

\begin{equation}
  \ln \mathcal{L} = - \displaystyle \ln (c\,\tau)\ \sum_i w_i' - 1/(c\,\tau) \sum_i \ w_i'\cdot l_i/(\beta_i\gamma_i) 
\end{equation}

in which the coefficient $w_i'$ corresponds to the normalized efficiency correction and the subtraction factor in case of the sideband data set. Additionally, the weight factor is normalized in order to keep the same integral count $N$ after applying the efficiency correction $f_i$ to the data set \cite{cite:james2006statistical}. Under these conditions, $w_i' = w_i \cdot N/(\sum_i w_i)$ with $w_i =  - N_b/f_i$ and $w_i = 1/f_i$ for the sideband and signal data sets respectively. The interval estimation at standard deviation of $\pm 1\sigma$ (68.3 \% confidence level) is obtained by the profiled likelihood ratio method \cite{cite:roostats}. The profiled likelihood ratios for estimation of the mean decay length $c\tau$ of $\Lambda$ hyperon, $^3_{\Lambda}$H and $^4_{\Lambda}$H hypernuclei are shown in the top panels of Figure \ref{fig:profile}. Deduced lifetime values for $\Lambda$, $^3_\Lambda $H and $^4_\Lambda $H are respectively $262 ^{+56}_{-43}$ ps, $183 ^{+42}_{-32}$ ps and $140 ^{+48}_{-33}$ ps.

In order to represent these values, the {\it binned} data of decay length $l/\beta\gamma$ in the signal region is displayed with the fitted model, which included the binned contribution of the exponential probability density function added to the {\it binned} data of the sideband regions. The binned fitted model at the bin $i$ can be expressed like this: 

\begin{equation}
\mathrm{ModelFit}_i = \left\lbrace \displaystyle \int_{bin_i} A/(c \tau) \exp(-l/(\beta\gamma\,c \tau))dl\right\rbrace  \cdot 1/w_i' +\ \mathrm{B_l}[ l_i/(\beta_i\gamma_i)]\ +\ \mathrm{B_h}[l_i/(\beta_i\gamma_i)] 
\end{equation}

where $A$ is the normalization factor of the exponential probability density function, $\mathrm{B_l}[l_i/(\beta_i\gamma_i)]$ and $\mathrm{B_h}[l_i/(\beta_i\gamma_i)]$ are the two sideband data sets at bin $i$.

The latter are shown in the bottom panels of Figure \ref{fig:profile}. Inference and interval estimation, via the maximum likelihood method, do not provide any information about the goodness-of-fit resulting from the inference. However, it is important to demonstrate that the estimation of the lifetimes describes the data set of the experimental observations. The goodness-of-fit is then tested by calculating the reduced $\chi^2$ of the model proper decay time function over the {\it binned} data as $\chi^2 = \sum_{bin_i} (\mathrm{SB}[l_i/(\beta_i\gamma_i)]-\mathrm{ModelFit}_i)^2/(\mathrm{SB}[l_i/(\beta_i\gamma_i)])$, as shown in the same panels. The values of reduced $\chi^2$ are 0.43 for $\Lambda$, 1.14 for $^3_\Lambda $H and 0.71 for $^4_\Lambda $H, respectively. It should be emphasized once more here that the deduced lifetime values have been obtained by the {\it unbinned} fitting, and the reduced $\chi^2$ from the binned representation was used only to cross check the goodness-of-fit.

\subsection{Study of the systematic uncertainty}

\begin{table*}[tHb] %
\centering
\caption{\label{tab:systematic} Contributions to the systematic errors. }
\begin{tabular}{cccc}
\toprule
Contribution to the systematic & $\Lambda$  & $^3_{\Lambda}$H & $^4_{\Lambda}$H \\
\midrule

Vertex Z pos (\%) & 8  & 18 & 14 \\
Primary Vertex (\%) & 4 & 4 & 8 \\
Scaling (\%)     & 11 & 6  & 17 \\
Sideband (\%)  & 8  & 6  & 10 \\
Total (\%)       & 17 & 20 & 25 \\

 \bottomrule
\end{tabular}
\end{table*}

The systematic uncertainty in deducing the lifetime was also investigated, as summarized in Table \ref{tab:systematic}. First, the boundary range of the longitudinal vertex position cut was set to several values in order to deduce the effect of the cut condition. The resultant systematic uncertainty is listed in the second row of Table~\ref{tab:systematic} (Vertex Z pos). It is mainly due to the change in the data sample's size, which influences the fitting procedure. 
In the lifetime estimation, the primary vertex position is not perfectly defined. Variations on the primary vertex can be studied to determine possible systematic influence to lifetime uncertainty. As explained the primary vertex position is restrained by the beam hit position on the TOF-start detector, however it does not define well enough the vertex position in 3-dimensions. By means of Monte Carlo simulations for the missing information on the primary vertex position, the systematic uncertainty on the primary vertex is extracted and listed in the third row (Primary Vertex). The effect of the scaling factor between the background contribution from the sidebands and the signal region was examined to measure its share of the systematic error. Those values are listed in the fourth row (Scaling). The sideband regions for the background contribution were also alternated to different values, which yielded another few percentage points. These contributions are listed in the fifth row (Sideband). The total systematic errors are 17 \% for $\Lambda$, 20 \% for $^3_{\Lambda}$H and 25 \% for $^4_{\Lambda}$H.

\begin{table*}[tHb] %
\centering
\caption{\label{tab:bias} Estimate of the bias of the lifetime estimation. }
\begin{tabular}{cccc}
\toprule
 & $\Lambda$  & $^3_{\Lambda}$H & $^4_{\Lambda}$H \\
\midrule
Bias estimation (ps) & 3.7  & -3 & -5 \\
Bias estimation (\%) & 1.4  & -1.6 & -3 \\
\bottomrule
\end{tabular}
\end{table*}

It is important to determine the bias of the lifetime estimator via the unbinned maximum likelihood method in order to validate the various deduced values. Since the \emph{true} probability density function of the full data set is unknown, the bias of the estimator can not simply be computed. However an estimate of the bias can be arrived at via bootstrapping \cite{cite:james2006statistical,efron1993introduction}. This is done by creating two bootstrap samples of the full dataset, which are a two disjoint sub-assembly $s1$, $s2$ of the full data set $s_{all}$. Then the full statistical analysis of the lifetime estimation is performed over those sub-assemblies. The bias estimator hence is then defined by~:~$Bias[\hat{\tau}] = 1/2\cdot(\hat{\tau}_{s1}+\hat{\tau}_{s2}) - \hat{\tau}_s$ where the operator $\hat{.}$ corresponds to the lifetime estimation by the unbinned maximum likelihood method \cite{cite:james2006statistical,efron1993introduction}. The estimation of bias of the lifetime estimator is summarized in Table \ref{tab:bias} and shows that the maximum likelihood method introduced a very small bias into the estimation of the lifetime of $\Lambda$, $^3_\Lambda $H and $^4_\Lambda $H.

\subsection{Comparison with previous experiments}

\begin{figure*}[thb]
\centering
    \begin{tabular}{cc}
   \resizebox{55mm}{!}{\includegraphics{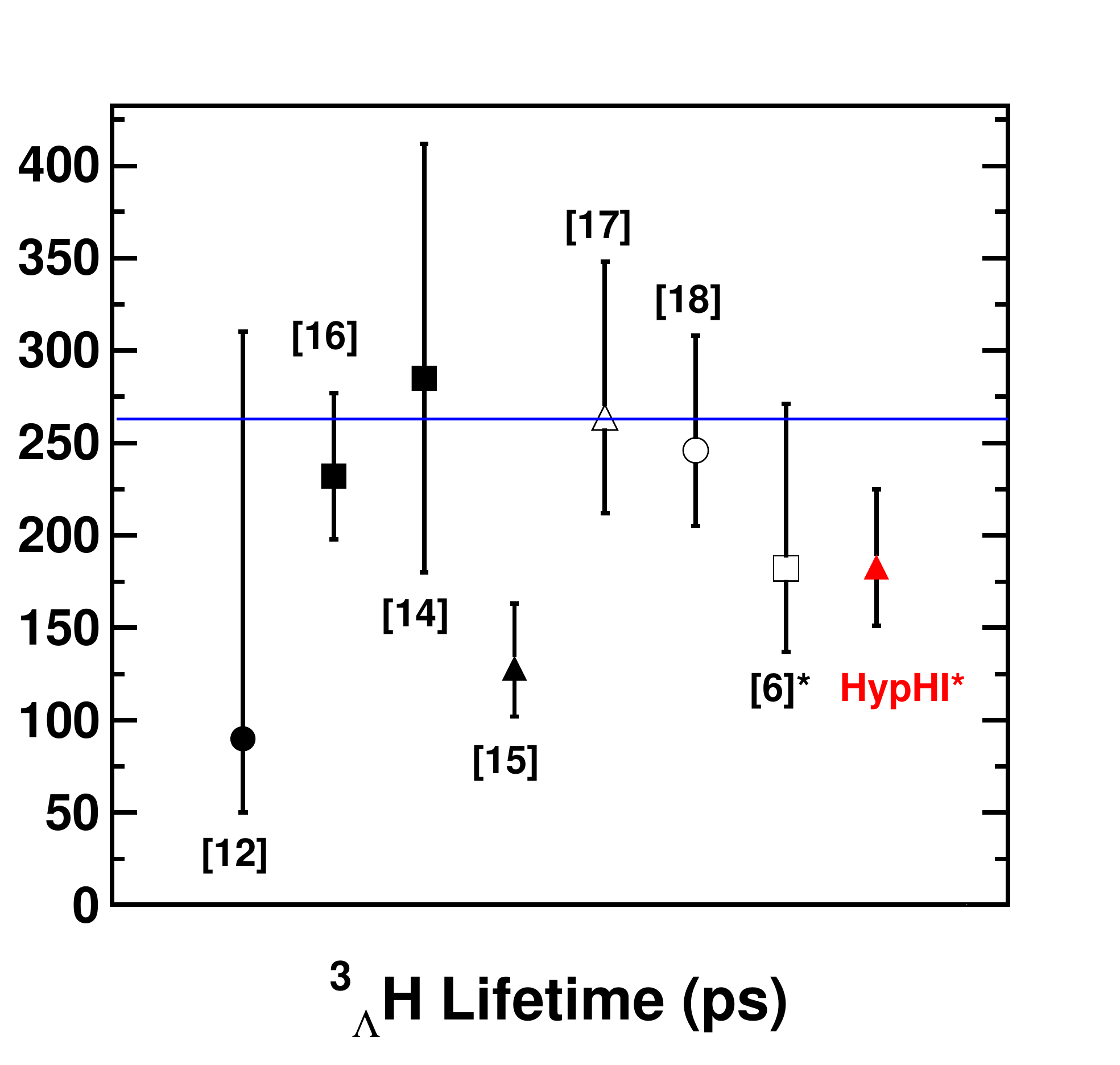}} & 
   \resizebox{55mm}{!}{\includegraphics{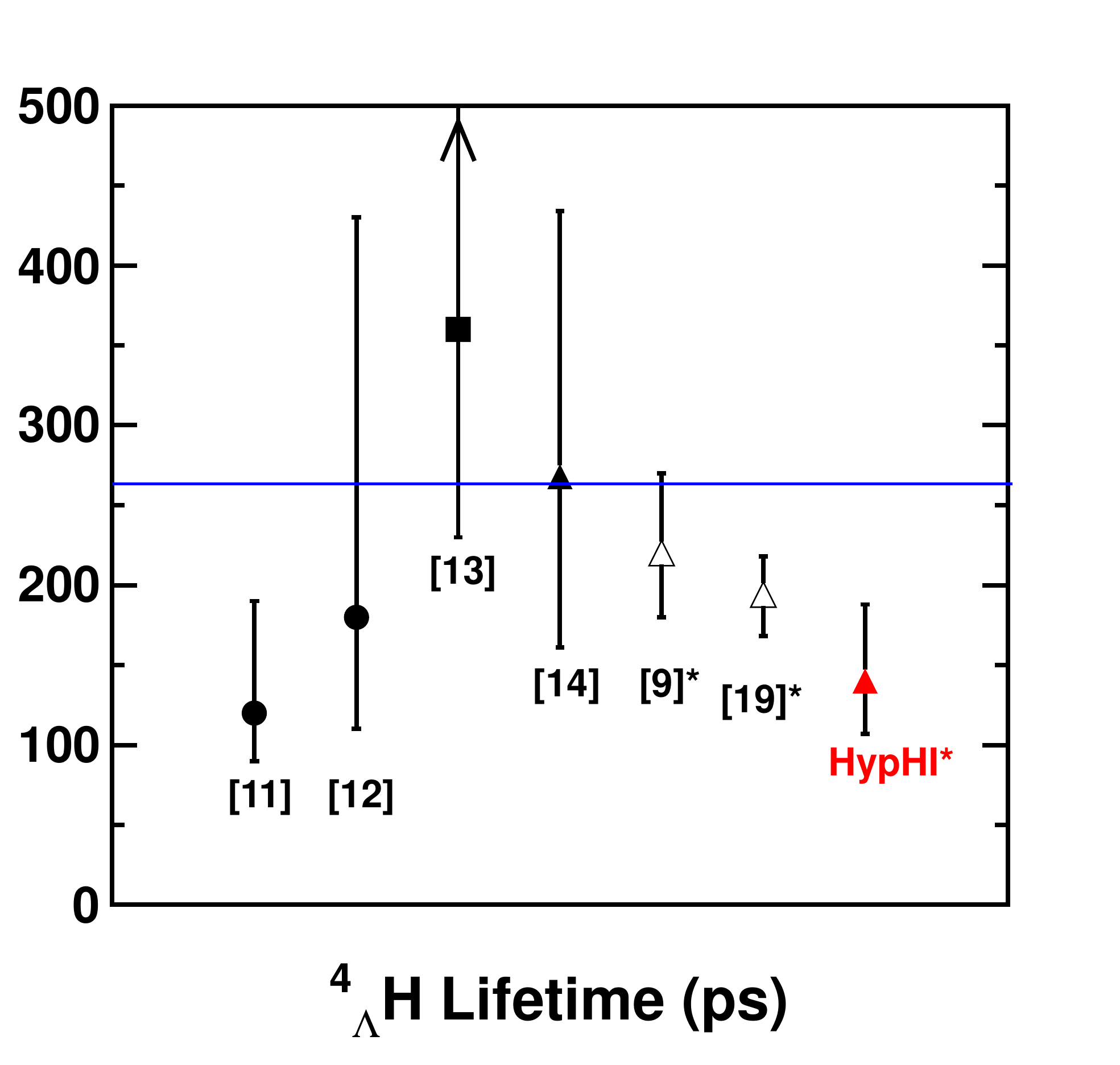}} \\ 
    \end{tabular}

   \caption{\label{fig:comp}(Color online) World data comparison of $^3_{\Lambda}$H and $^4_{\Lambda}$H lifetimes. Values deduced in this experiment are indicated by "HypHI". The horizontal line at 263.2 ps shows the known lifetime of the $\Lambda$ hyperon. References to counter experiments are marked by an asterisk.}
         
\end{figure*}

The measured value of the lifetime of the $\Lambda$ hyperon compares favorably with the well known value of 263.2 $\pm$ 2.0 ps \cite{cite:pdg}. A comparison of the deduced lifetime values of $^3_{\Lambda}$H and $^4_{\Lambda}$H hypernuclei with previously published values is shown in Figure \ref{fig:comp}. It can be stated that our measured values for $^3_{\Lambda}$H and $^4_{\Lambda}$H hypernuclei are smaller than the lifetime of $\Lambda$ hyperon of 263.2 $\pm$ 2.0 ps.

\section{Conclusion}

In summary, the HypHI Phase 0 experiment was performed by using a heavy ion-induced reaction with $^6$Li projectiles at 2 $A$ GeV on a carbon target. The invariant mass distributions of the final states of p+$\pi ^-$, $^3$He+$\pi ^-$ and $^4$He+$\pi ^-$ revealed signals of $\Lambda $, $^3_{\Lambda}$H and $^4_{\Lambda}$H, respectively with significance values of 6.7$\sigma$, 4.7$\sigma$ and 4.9$\sigma $. The lifetime values were deduced by using unbinned maximum likelihood fitting, and they are $262 ^{+56}_{-43} \pm 45 $ ps for $\Lambda $, $183 ^{+42}_{-32} \pm 37 $ ps for $^3_{\Lambda}$H and $140 ^{+48}_{-33} \pm 35$ ps for $^4_{\Lambda}$H. The current work has demonstrated the feasibility of measuring the mass and the lifetime of various hypernuclei with induced reactions of heavy ion beams on a fixed target.

\section{Acknowledgments}

The authors would like to thank the GSI Departments of Accelerator, of Experimental Electronics, of the Detector Laboratory and of the Target Laboratory and the Electronics Department of the Institute for Nuclear Physics of Mainz University for supporting the project. The HypHI project is funded by the Helmholtz association as Helmholtz-University Young Investigators Group VH-NG-239 at GSI, and the German Research Foundation (DFG) under contract number SA 1696/1-1. The authors acknowledge the financial support provided by the Ministry of Education, Science and Culture of Japan, Grant-in-Aid for Scientific Research on Priority Areas 449, and Grant-in-Aid for promotion of Cooperative Research in Osaka Electro-Communication University (2004-2006). This work is also supported by the Ministry of Education, Science and Culture of Japan, Grants-in-Aid for Scientific Research 18042008 and EU FP7 Hadron-Physics-2 SPHERE. A part of this work was carried out on the HIMSTER high performance computing infrastructure provided by the Helmholtz-Institute Mainz.

\bibliography{biblio}

\end{document}